\documentclass[twocol]{ametsocV6.1}

\usepackage{sah_shortcuts}

\title{Interpreting seasonal and interannual Hadley cell descending edge migrations via the cell-mean Rossby number}

\authors{Spencer A. Hill,\aff{a}\correspondingauthor{Spencer A. Hill, shill1@ccny.cuny.edu} 
Simona Bordoni,\aff{b} 
Jonathan L. Mitchell,\aff{c,d} 
Juan M. Lora,\aff{e}
}

\affiliation{\aff{a}{Department of Earth and Atmospheric Sciences, City College of New York}\\
\aff{b}{Department of Civil, Environmental and Mechanical Engineering (DICAM), University of Trento, Italy}\\
\aff{c}{Department of Earth, Planetary, and Space Sciences, University of California, Los Angeles}\\
\aff{d}{Department of Atmospheric and Oceanic Sciences, University of California, Los Angeles}\\
\aff{e}{Department of Earth \& Planetary Sciences, Yale University, New Haven, Connecticut}\\
}

\abstract{%
The poleward extent of Earth's zonal-mean Hadley cells varies across seasons and years, which would be nice to capture in a simple theory.  A plausible, albeit diagnostic, candidate from \citet{hill_theory_2022} combines the conventional two-layer, quasi-geostrophic, baroclinic instability-based framework with a less conventional assumption: that each cell's upper-branch zonal winds are suitably captured by a single, cell-wide Rossby number, with meridional variations in the local Rossby number neglected.  We test this theory against ERA5 reanalysis data, finding that it captures both seasonal and interannual variations in the Hadley cell zonal winds and poleward extent fairly well.  For the seasonal cycle of the NH cell poleward edge only, this requires empirically lagging the prediction by one month, for reasons unclear to us.  In all cases, the bulk Rossby number value that yields the most accurate zonal wind fields is approximately equal to the actual, diagnosed cell-mean value.  Variations in these cell-mean Rossby numbers, in turn, predominantly drive variations in each cell's poleward extent.  All other terms matter much less---including the subtropical static stability, which, by increasing under global warming, is generally considered the predominant driver of future Hadley cell expansion.  These results argue for developing a predictive theory for the cell-mean Rossby number and for diagnosing its role in climate model projections of future Hadley cell expansion. 
} 

\begin{document}
\maketitle

\section{Introduction}

Descent in the poleward branches of the time-mean, zonal-mean Hadley cells promotes aridity in the subtropics, fundamentally shaping Earth's hydrological cycle.  How far this descent spans poleward, and why, motivated foundational studies of Earth's general circulation
\citep{halley_historical_1686,hadley_concerning_1735,ferrel_essay_1856} and remains actively investigated.  In both hemispheres it contracts equatorward in winter to spring and expands poleward in summer to autumn; interannually it tends to contract equatorward in the warm, El Ni\~no phase of the El Ni\~no Southern Oscillation (ENSO) and to expand poleward in the cool, \lanina/ phase, though more reliably for the Southern Hemisphere (SH) than the Northern Hemisphere (NH) \citep{caballero_role_2007,lu_response_2008,tandon_understanding_2013,zurita-gotor_coupled_2018,seo_what_2023}.  Here, we show that a simple theory captures these seasonal and interannual migrations in reanalysis data fairly well and is likely applicable in other contexts, such as the cells' poleward expansion under global warming \citep[\eg/][]{vallis_response_2015,chemke_exploiting_2019,staten_tropical_2020}.

Dynamically, each hemisphere's poleward Hadley cell edge demarcates the transition between the extratropics, which are strongly influenced by large-scale baroclinic eddies, and the tropics where those baroclinic eddies are much less prevalent \citep{showman_atmospheric_2014}---making it natural to link the edge's location to where baroclinic instability first sets in \citep{held_general_2000,kang_expansion_2012,hill_theory_2022,peles_estimating_2023}.  In the influential model introduced by \citet{held_general_2000}, this is determined by where the upper-layer westerlies of the Hadley cells exceed the critical speed for baroclinic instability according to a simple, two-layer, quasi-geostrophic model.  On a given planet with fixed gravitational constant, planetary radius, and planetary rotation rate, the two terms that can change this critical zonal wind field are the tropopause height and the static stability, and prior studies have affirmed the predominance of the subtropical static stability in warming-induced Hadley cell expansion \citep[\eg/][]{chemke_exploiting_2019}.

For given values of these two parameters and thus of the critical zonal wind field, the onset latitude still depends on the Hadley cell upper-branch zonal wind field, which \citet{held_general_2000} took as the angular momentum conserving zonal winds with ascent at the equator \citep{held_nonlinear_1980}.  But the same baroclinic eddies, by propagating equatorward and breaking within the Hadley cells, decelerate winds well below this limit \citep{walker_response_2005,walker_eddy_2006,schneider_general_2006,sobel_single-layer_2009}.  So too do migrations of the Hadley cell \emph{ascending} edge off the equator, by reducing the planetary angular momentum value imparted to the free troposphere \citep{kang_expansion_2012,hilgenbrink_response_2018,watt-meyer_itcz_2019}, as occurs seasonally when the ascending branch expands into either summer hemisphere \citep{lindzen_hadley_1988}.  The decelerations by either mechanism, in turn, move the baroclinic instability onset latitude poleward.
These factors---a descending edge determined by baroclinic instability onset and Hadley cell zonal wind fields that depend on both the aggregate extratropical eddy stresses and ascending edge migrations---have been formalized by \citet{hill_theory_2022}, drawing heavily from \citet{kang_expansion_2012}.

A central ansatz of the \citet{hill_theory_2022} theory is that the local Rossby number, a scalar formally defined below that essentially quantifies the strength of the eddy stresses at each latitude, is uniform within each Hadley cell's upper branch at any given time, and consequently that each Hadley cell's upper-branch zonal winds are suitably approximated by those required by a uniform Rossby number.  Crucially, this cell-mean Rossby number value can differ between the NH and SH cells at any particular time and for either cell can vary in time, such as seasonally or interannually \citep[\eg/][]{schneider_general_2006,bordoni_monsoons_2008,kang_expansion_2012}.  Our purpose here is to test that ansatz and the theory's broader fidelity for Earth's current climate, as represented in reanalysis data.  After formally presenting the theory (Section~\ref{sec:theory}) and methodological choices (Section~\ref{sec:methods}),
we show that, unsurprisingly \citep[\eg/][]{schneider_general_2006,caballero_role_2007,singh_limits_2019}, the Rossby number is far from uniform within either Hadley cell almost always, but that, more surprisingly, the uniform-Rossby number zonal wind fields nevertheless capture the actual wind fields quite well, both for the climatological seasonal cycle (Section~\ref{sec:seasonal}) and interannual variability (Section~\ref{sec:interann}).  Moreover, the theory captures the Hadley cell descending edges fairly well in both contexts.

Several prior studies affirm the central role in reanalysis data of variations in eddy stresses on variations of the Hadley cells interannually \citep{caballero_role_2007,zurita-gotor_coupled_2018,seo_what_2023}.  More recently, \citet{peles_estimating_2023} have done so for the climatological seasonal cycle, showing for both hemispheres in JRA-55 data that multiple bulk measures of the equatorward extent of baroclinic instability track the climatological seasonal cycle of the Hadley cell well.  In effect, the present study seeks to further distill these prior results into the simplest plausible analytical framework, and then use that framework to evaluate the relative importance of each term appearing---the cell-mean Rossby number, the ascending edge latitude, the static stability, and the tropopause height---finding the cell-mean Rossby number to predominantly control the Hadley cell descending edge migrations both seasonally and interannually.  We conclude by discussing some implications of these results, including for forced Hadley cell expansion under global warming (Section~\ref{sec:conc}).

\section{Theory}
\label{sec:theory}
The theory was originally presented by \citet{hill_theory_2022} and draws heavily from \citet{kang_expansion_2012}.  The local Rossby number is
\begin{equation}
  \label{eq:ross-num}
  \Ro(\lat)\equiv-\frac{\zeta}{f},
\end{equation}
where \(\lat\) is latitude, \(\zeta\equiv(a\cos\lat)^{-1}\partial_\lat(u\cos\lat)\) is the zonal-mean relative vorticity with zonal-mean zonal wind \(u\), and \(f\equiv2\Omega\sin\lat\) is planetary vorticity.  In what follows, \(u\) and all other fields are taken to be monthly or longer averages.

If the local Rossby number is meridionally uniform, \(\Ro(\lat)\equiv\Ro\), then integrating (\ref{eq:ross-num}) meridionally yields the uniform-Rossby number zonal wind field
\begin{equation}
  \label{eq:uro}
  \uro(\lat;\Ro,\lata)=\Ro\,\uamc(\lat;\lata)=\Ro\,\Omega a\left(\frac{\sin^2\lat-\sin^2\lata}{\cos\lat}\right),
\end{equation}
where \(\uamc\) is the angular momentum conserving zonal wind field given the ascending edge \(\lata\) \citep{held_nonlinear_1980,lindzen_hadley_1988}.  As ascent moves away from the equator, a smaller planetary vorticity value is advected into the free troposphere, causing the resulting uniform-\(\Ro\) zonal winds to be less westerly at all latitudes.  The angular momentum conserving solution amounts to the limiting case of \(\Ro=1\).   

Meanwhile, the critical upper-level zonal wind for baroclinic instability in the two-layer quasi-geostrophic model is given by 
\begin{equation}
  \label{eq:ubci}
  u_\mr{BCI}(\lat;H,\deltav)=\frac{gH\deltav}{2\Omega a}\frac{\cos\lat}{\sin^2\lat},
\end{equation}
where \(g\) is gravity, \(H\) is the tropopause height, \(\Omega\) is the planetary rotation rate, \(a\) is the planetary radius, and \(\deltav\) is a dimensionless bulk static stability.\footnote{Note that \citet{hill_theory_2022} mistakenly omit the factor of 2 in the denominator of (\ref{eq:ubci}).  Fortunately this does little damage to the resulting prediction for the descending edge, because the missing \(2^{-1/4}\approx0.84\) constant factor that results in the descending edge prediction can implicitly be subsumed into the empirical fitting constant \(c_\mr{d}\), whose precise value is already not taken seriously.}  More precisely, this is an expression for the critical vertical shear in the zonal wind, but we assume as standard that the lower-layer zonal wind magnitude is negligible \citep{held_general_2000}.

\begin{figure*}[h]
  \centering
  \includegraphics[width=0.55\textwidth]{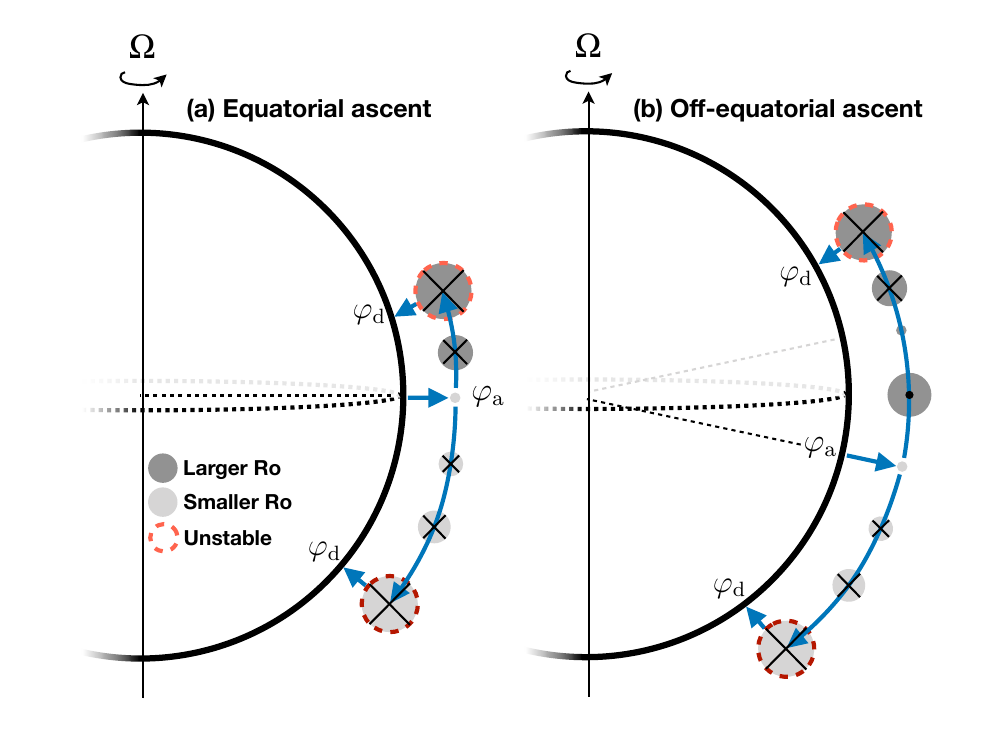}
  \includegraphics[width=0.44\textwidth]{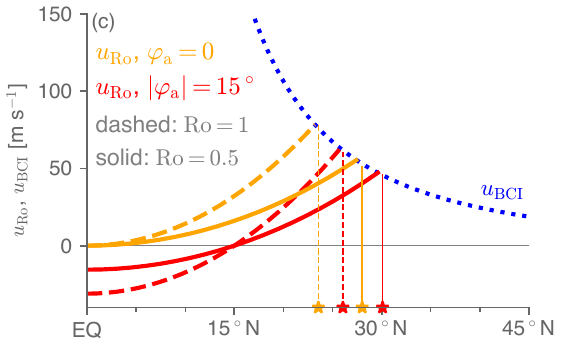}\\
  \caption{Schematic of the influences of the mean upper-tropospheric Rossby number in each Hadley cell, $\Ro$, as well as the ascent latitude, $\lata$, on the latitude of the descending edge, $\latd$, according to our theory.  In (a) and (b), blue arrows signify the locations of ascent, descent, and the directions of the meridional flow in each cell's upper branch, and the cell-mean Rossby number is smaller for the Southern Hemisphere cell compared to the Northern Hemisphere cell, enabling the Southern Hemisphere cell to extend farther poleward before its zonal winds (grey circles with black Xs for westerlies or dots for easterlies, with larger symbols indicating larger magnitudes) become baroclinically unstable, thereby causing the cell to terminate.  In (b), ascent is off equator, which compared to the corresponding on-equatorial ascent case in the left panel results in less westerly zonal winds and thus further poleward Hadley cell extents.  (c) Uniform-Rossby number zonal wind fields, \(\uro\), with either (orange) equatorial ascent or (red) ascent at 15\(^\circ\)N or S, each with either (dashed) \(\Ro=1\) or (solid) \(\Ro=0.5\).  The dotted blue line is \(u_\mr{BCI}\), the critical zonal wind for baroclinic instability, using \(\deltav=1/8\), \(H=10\)~km, and standard values of all other constants.  Stars and vertical lines at the intersections of the \(\uro\) fields with \(u_\mr{BCI}\) signify the prediction for the descending edge \(\latd\) for that \(\uro\).}
  \label{fig:schematic}
\end{figure*}

We then solve for the latitude where instability first sets in by setting \(\uro=u_\mr{BCI}\).  Equating the right-hand sides of (\ref{eq:uro}) and (\ref{eq:ubci}), the result in the small-angle limit is
\begin{equation}
  \label{eq:bci-edge-off-eq}
  \latd^2=c_\mr{d}^2\left(\frac{\lata^2}{2}+\sqrt{\frac{\lata^4}{4}+\frac{\Bu\deltav}{2\Ro}}\right),
\end{equation}
where
\begin{equation}
  \label{eq:burg}
  \Bu\equiv \frac{gH}{(\Omega a)^2}
\end{equation}
is the planetary Burger number, and we have also introduced a near-unity empirical fitting constant, \(c_\mr{d}\): denoting the true solution as \({\lat_\mr{BCI}}\), we set \({\latd\equiv c_\mr{d}\lat_\mr{BCI}}\).  In (\ref{eq:bci-edge-off-eq}), the Rossby number is what encapsulates the deceleration of the Hadley cell zonal winds by extratropical eddy stresses.  We will show that it can be interpreted as a meridional average of the local Rossby number within the upper branch of each Hadley cell.
From (\ref{eq:bci-edge-off-eq}), the descending edge moves poleward as \(\Ro\) decreases or \(\lata\) moves poleward.  
Only the absolute value of the ascending edge, \(|\lata|\) rather than \(\lata\), influences the descending edge prediction, a consequence of the symmetry about the equator of \(\uro\): \({\uro(\lat;\Ro,\lata)=\uro(|\lat|;\Ro,|\lata|)}\).  In other words, whether the ascent is in the northern or southern hemisphere does not matter; what matters is how far poleward it is into either hemisphere.  These behaviors are shown schematically in Fig.~\ref{fig:schematic}.  Absent from Fig.~\ref{fig:schematic} are the influences of changes in \(H\) and \(\deltav\).  Both move the descending edge prediction poleward as they increase, all else equal,
but as shown below they play negligible roles in the seasonal cycle and interannual variability.

Though \citet{hill_theory_2022} combine (\ref{eq:bci-edge-off-eq}) with a separate theory that predicts the ascending edge latitude based on the extent of tropical supercriticality \citep{hill_solsticial_2021}, we find the abrupt jumps of the ascending edge over the seasonal cycle predicted therein to be inconsistent with the more nearly sinusoidal seasonality in the ascending edge in reanalysis data shown below \citep{dima_seasonality_2003}---which is likely strongly influenced by zonally asymmetric eddies \citep{walker_response_2005,schneider_eddy-mediated_2008,geen_processes_2019} in addition to thermal inertia \citep{wei_energetic_2018,zhou_hierarchy_2018}.  We therefore proceed using (\ref{eq:bci-edge-off-eq}), and emphasize that this is a diagnostic expression, lacking as we do compelling predictive theories for either the ascending edge latitude or the cell-mean Rossby number.

In the case of on-equatorial ascent, \(\lata=0\), (\ref{eq:bci-edge-off-eq}) simplifies to
\begin{equation}
  \label{eq:bci-edge-eq}
  \lat_{\mr{d},0}\equiv\latd({\lata=0})=c_\mr{d}\left(\frac{\Bu\deltav}{2\Ro}\right)^{1/4},
\end{equation}
from which one sees that, once the \(c_\mr{d}\) constant is distributed through, the last term in (\ref{eq:bci-edge-off-eq}) is equivalent to \(\lat_{\mr{d},0}^4\).  The expression (\ref{eq:bci-edge-eq}) reduces to the original \citet{held_general_2000} expression in the \(\Ro=1\), angular momentum-conserving limit.

As \citet{held_general_2000} notes, some of the two-layer quasi-geostrophic assumptions are tenuous when applied to the real atmosphere, and more recently \citet{peles_estimating_2023} nicely summarize and expand upon the class of approaches that incorporate more comprehensive physics.  In particular,
in continuously stratified atmospheres no critical shear exists that determines baroclinic instability onset, and prior studies  \citep{korty_extent_2008,levine_response_2011,levine_baroclinic_2015} have instead framed the eddy influence on the cell edges in terms of extratropical supercriticality, a measure of the vertical extent of eddy heat fluxes within the troposphere \citep{held_vertical_1978}.  Also, arguably the Hadley cell descending edges depend less directly on the location of wave generation, which the instability criterion nominally corresponds to, than on the location of wave breaking \citep{vallis_response_2015}, which in turn is not even a single latitude, varying as it does with the eddy phase speed \citep[\eg/][]{chen_phase_2007}.  Finally, despite considerable theoretical, modeling, and reanalysis-based evidence that extratropical eddy processes do fundamentally control the Hadley cell descending edge latitudes, the causality has recently been questioned, at least in the context of warming-induced poleward expansion \citep{davis_eddy-mediated_2022}.  The reader should keep these important caveats in mind as we proceed in using the simple metric (\ref{eq:bci-edge-off-eq}) to predict $\latd$.

\section{Methods}
\label{sec:methods}
\subsection{Reanalysis data}
We use monthly fields from the European Center for Medium-range Weather Forecasting (ECMWF) ERA5 reanalysis dataset \citep{hersbach_era5_2020} spanning January 1979 to December 2023.  These are provided by ECMWF transformed from the underlying numerical model's native coordinate system to a 0.25$\times$0.25$^\circ$ latitude-longitude horizontal grid and to 37 fixed pressure levels from 1000 hPa to 1 hPa in the vertical.  We compute annual, seasonal, and calendar-month climatologies over 1979-2023, as well as timeseries of fields averaged over each calendar year.

\subsection{Standard Hadley cell diagnostics}
As standard, the starting point for our Hadley cell diagnostics is the Eulerian-mean meridional mass overturning streamfunction, $\Psi$:
\begin{equation}
  \label{eq:streamfunc}
  \Psi(\lat,p)=2\pi a\cos\lat\int_0^pv\,\frac{\mr{d}p'}{g},
\end{equation}
where $a$ is planetary radius, $\lat$ is latitude, $p$ is pressure, $v$ is zonal-mean meridional wind, and $g$ is gravitational acceleration.  It is signed such that the SH Hadley cell is negative and the NH cell is positive.  Though not shown in (\ref{eq:streamfunc}), as standard at each latitude the column average of $v$ is subtracted from $v$ at all levels barotropically to enforce vanishing column-integrated mass transport and thus that $\Psi$ vanishes at the upper and lower boundaries.

Rather than the conventional zero-crossing of the 500-hPa streamfunction \citep[\eg/][]{adam_tropd_2018}, the descending edge latitudes are diagnosed as where the streamfunction vertically averaged over 500-800 hPa decreases to 5\% of its value at the center, linearly interpolating between the two grid points bracketing this threshold crossing.  The ascending edge is defined in the same way, moving northward from the SH cell center, and the cell center is defined as the latitude and pressure where the streamfunction takes its maximum magnitude within a given Hadley cell.

\begin{figure*}[h]
  \centering
  \includegraphics[width=\textwidth]{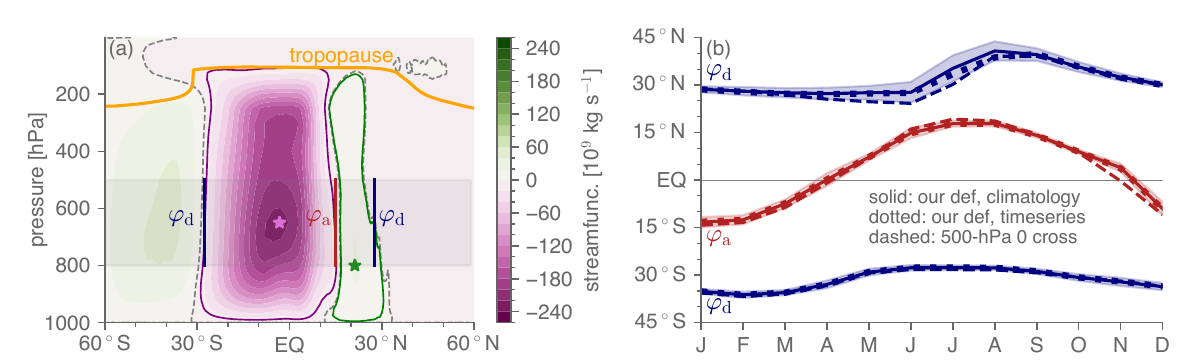}
  \caption{(a) For the June climatology, meridional mass streamfunction in color shading according to the colorbar.  The 500-800 hPa layer is shaded grey, and our diagnosed cell edges are overlaid over that span at their latitudes.  Pink and green stars show the SH and NH cell center locations, respectively.  Solid purple and green curves are 5\% of the SH and NH Hadley cell maxima, respectively, and dashed grey curves are the streamfunction zero crossings.  The orange curve is the local tropopause.  (b) Diagnosed monthly climatological Hadley cell edges.  Solid lines use the streamfunction for each calendar month climatology averaged over 500-800 and using a 5\% threshold from the center.  Dotted lines are the same but applied to each individual month and then averaged across years.  Dashed lines are the conventional 500-hPa zero crossing applied to the monthly climatologies.  Shading shows the \(\pm1\) interannual standard deviation range for our 500-800-hPa based metric.}
  \label{fig:edge-sens}
\end{figure*}

Fig.~\ref{fig:edge-sens}a illustrates the factors motivating these choices using the month of June.  An average over the 500-800 hPa layer (grey shading) is chosen as a compromise between the conventional 500~hPa level and the fact that, except for the NH cell in April and May, the cell maximum strengths all occur at or vertically below 600~hPa (pink and green stars for June; not shown for other months).  The choice of a 5\% threshold rather than a true zero crossing is primarily motivated by the NH cell being very weak during the boreal summer months \citep[\eg/][]{watt-meyer_hemispheric_2019}: the SH cell strength exceeds that of the NH by more than tenfold in each of the June, July, and August calendar month climatologies, the JJA seasonal-mean climatology, and in multiple boreal-summer months in 36 of the 45 years (not shown).  As such, in some individual boreal summer months, the NH Hadley cell is so weak and poorly organized that the SH cell, rather than passing sharply through zero, gradually tails off moving northward, making zero-crossing definitions of both the ascending and NH descending edge unphysical.

As one additional consideration---though admittedly not one directly relevant to our results---the 5\% threshold constitutes a meaningful boundary of each Hadley cell in all directions, as shown by the overlaid solid purple and green contours in Fig.~\ref{fig:edge-sens}a, which is not true of the zero crossing shown in dashed grey contours.  In particular, the 5\% threshold for the SH cell is nearly coincident with the tropopause (overlaid orange curve, defined further below), consistent with the Hadley cells being confined to the troposphere.  

Fig.~\ref{fig:edge-sens}b shows the climatological seasonal cycles of all three edges and illustrates their metric and sampling sensitivities.  Solid curves use our methods computed from the climatological streamfunction for each calendar month.  Dotted curves are the same, but computed for each individual month and then averaged for each calendar month across years.  Because locating the cell center and threshold crossings are not linear operations, in principle this choice of order of operations could matter \citep{adam_tropd_2018}, but in practice it makes little difference here.  More substantive differences, though still modest overall, arise from using the conventional 500-hPa zero crossing metric, shown in dashed curves.  The ascending edge defined this way moves southward across the equator more rapidly in November-December, and the NH descending edge is more equatorward by a few degrees in boreal spring through summer.  From Fig.~\ref{fig:edge-sens}a for June at least, we see this stems from the bounding streamline of the NH cell tilting equatorward with height in the lower free troposphere.  Finally, shading about the solid curves in Fig.~\ref{fig:edge-sens}b shows the \(\pm1\) interannual standard deviation range for our 500-800-hPa based metric.  For reasons just discussed, it is largest in the NH during boreal summer, but still for all three edges it is modest relative to the seasonally forced climatological progression. 

\subsection{Uniform-\(\Ro\) framework diagnostics}

\begin{figure*}[h]
  \centering
  \includegraphics[width=\textwidth]{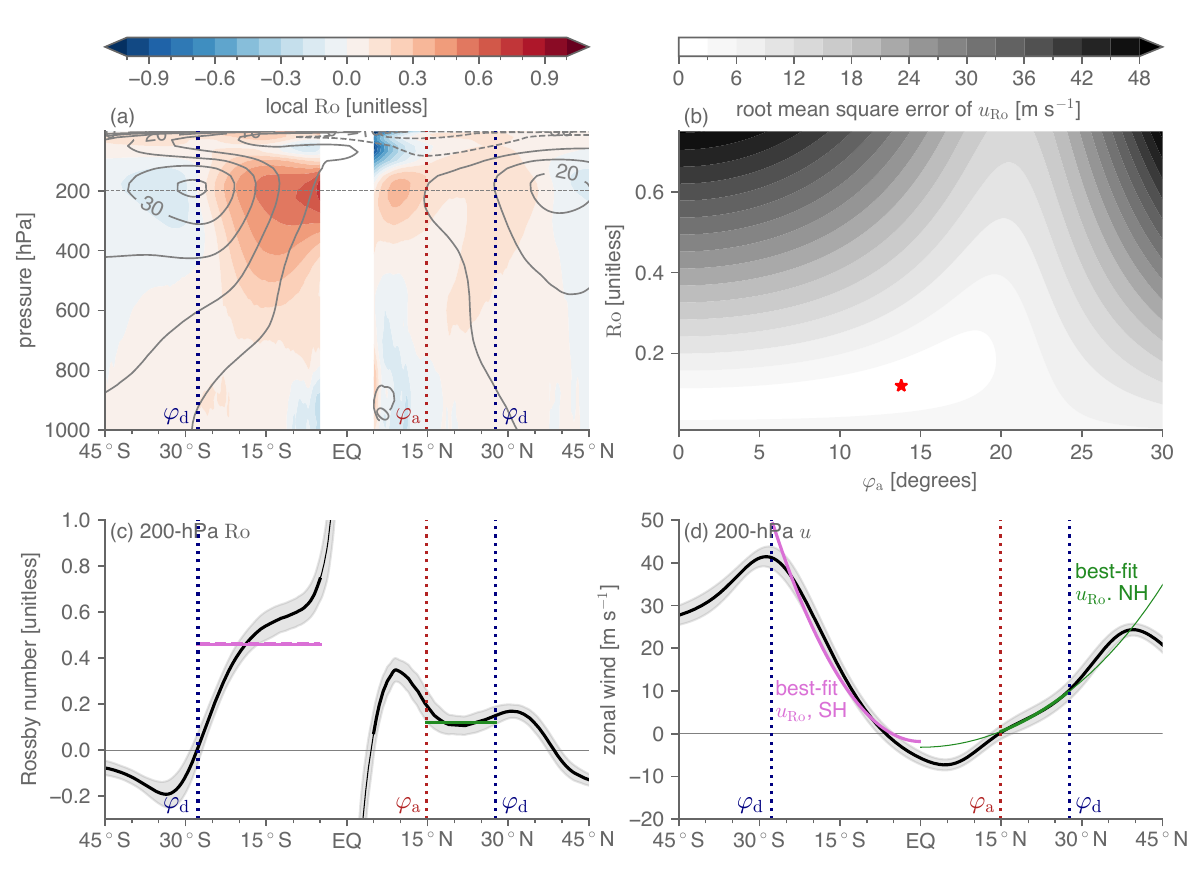}
  \caption{For the ERA5 1979-2023 June climatology, (a) local Rossby number at each latitude and pressure in color shading according to the color bar and masked withing 5\(^\circ\)S-5\(^\circ\)N, with local zonal wind overlaid in grey contours, with zero and positive values solid, negative dashed, and a 10~m~s\(^{-1}\) contour interval.  A thin horizontal line highlights the 200 hPa level.  (b) Root-mean-square-error of the uniform-Rossby number zonal wind field for the June NH cell as a function of the selected Rossby number and ascent latitude, with the red star indicating the minimum, from which the best-fit Rossby number and ascent latitudes are selected.  (c and d) Meridional profiles at 200-hPa of (c) local Rossby number and (d) zonal wind in black, with shading showing \(\pm\)1 interannual standard deviation.  In (a), (c), and (d), dotted vertical lines show the three diagnosed Hadley cell edge latitudes.  In (c), the thin black shows the values within 5\(^\circ\)S-5\(^\circ\)N where the Rossby number becomes ill-behaved, and pink and green horizontal lines show the (solid) diagnosed and (dashed) best-fit cell-mean Rossby numbers, with the meridional span indicating the region over which they are calculated.  In (d), pink and green curves show best-fit uniform-$\Ro$ zonal wind fields for the SH and NH Hadley cells, respectively.}
  \label{fig:fixed-ro-methods}
\end{figure*}

Fig.~\ref{fig:fixed-ro-methods} uses the results for the June climatology to illustrate our methodology for constructing the uniform Rossby number fields.  We start by computing the local Rossby number, \(\Ro(\lat)\), at each latitude and pressure using (\ref{eq:ross-num}), masking out values within 5\(^\circ\)S-5\(^\circ\)N where $\Ro(\lat)\propto 1/\sin\lat$ becomes ill-defined (Fig.~\ref{fig:fixed-ro-methods}a).

We then compute each Hadley cell's cell-mean Rossby number, $\Ro$, as an area-weighted meridional average of $\Ro(\lat)$ at 200 hPa, which is the tropospheric level where it invariably maximizes (Fig.~\ref{fig:fixed-ro-methods}a).  These span from that cell's diagnosed $\latd$ to whichever is closer: \(\lata\) or 5\(^\circ\) in the same hemisphere as \(\latd\), a choice again motivated by \(\Ro(\lat)\) being ill-defined near the equator  (Fig.~\ref{fig:fixed-ro-methods}c).  As such, for cross-equatorial cells this does not include the portion in the opposite hemisphere.

We fit $\uro$ to each Hadley cell via a two-dimensional parameter sweep over $\Ro$ and $\lata$, selecting the \((\Ro, \lata)\) pair that minimizes the root root mean square error (RMSE) of $\uro$ against the climatological zonal wind at 200~hPa, over the same meridional span as just described for the cell-mean Rossby number, except with no masking near the equator (Fig.~\ref{fig:fixed-ro-methods}b for the June NH cell; June SH cell and all others not shown).  For the climatological seasonal cycle, the sweep spans $0.01\leq\Ro\leq0.75$ in 0.01 increments and ${0\leq\lata\leq30^\circ}$ in 0.1$^\circ$ increments.  For interannual variability, where variations in both fields are smaller, the sweep spans \(0.2\leq\Ro\leq0.4\) in 0.002 increments and \(0\leq\lata\leq15^\circ\) in \(0.1^\circ\) increments.  The RMSE of the \(\uro\) fields is generally more sensitive to the specified Rossby number than the ascent latitude.

This procedure yields for each cell a best-fit \(\uro\) field (Fig.~\ref{fig:fixed-ro-methods}d) and corresponding best-fit values of $\Ro$ and $\lata$.  To interpret the physical significance of these four best-fit values, we compare them to their directly diagnosed counterparts: the two diagnosed cell-mean $\Ro$ values and the single diagnosed $\lata$ value as discussed further below.

We compute the tropopause height at each latitude, \(H(\lat)\), using the standard lapse rate-based definition as the lowest point where the local lapse rate drops below 2~K~km\(^{-1}\) \citep{wmo_meteorology_1957}, first refining the vertical resolution via cubic interpolation to 0.1~hPa-spaced levels.  The resulting \(H(\lat)\) field is very flat within the tropics, has a sharp gradient always poleward of the Hadley cell descending edges by a few degrees, and in the extratropics slopes more gradually down moving poleward.  We compute the bulk static stability at each latitude, \(\deltav(\lat)\), as the difference between the potential temperature at 500 and 850~hPa, divided by a standard reference value of 300~K: \(\deltav(\lat)\equiv(\theta_{500}(\lat)-\theta_{850}(\lat))/\theta_\mr{ref}\).  Climatologically, across all calendar months and non-polar latitudes this ranges over \({\sim}0.055-0.085\).  The specific levels of 500 and 850~hPa follow \citet{lu_response_2008}, motivated by eddy growth being more sensitive to lower- than upper-tropospheric baroclinicity.  As a sensitivity test, using 300 rather than 500~hPa as the upper level does not strongly influence our results (not shown).

In the two-layer, quasi-geostrophic model underlying the baroclinic instability onset criterion, the tropopause height and bulk static stability are fixed, global constants, and only by treating them as such can (\ref{eq:bci-edge-off-eq}) be derived.  But in the real atmosphere they vary nontrivially in space and time, and for the tropopause these variations are clearly strongly influenced by the extent of the Hadley cells, as can be inferred from Fig.~\ref{fig:edge-sens} for June.  We therefore create bulk mid-latitude values for each by meridionally averaging each over the 20\(^\circ\) latitude span of the extratropics beginning 5\(^\circ\) poleward of the Hadley cell descending edge.  As shown below, neither term strongly influences the Hadley cell descending edge predictions anyways, making this theoretical ambiguity not of practical importance.

We set the empirical fitting constant \(c_\mr{d}\) to a fixed value for each hemisphere, in each case chosen subjectively to provide, by eye, the best overall fit across all months for the climatological seasonal cycle or years for interannual variability.  For the seasonal cycle, these are 1.04 for the NH and 1.09 for the SH; for interannual variability they are 1.02 for the NH and 1.09 for the SH.  Given sampling and metric uncertainties---consider, for example, that the reference temperature of 300~K in the denominator of \(\deltav\) is somewhat arbitrary---and that the resulting values all end up being near unity, we do not consider the specific values of \(c_\mr{d}\) as being particularly meaningful; see \citet{hill_theory_2022} for further discussion.

For interannual variability, we compute yearly values of the NINO3.4 index over 1979-2023 using monthly data from the NOAA Extended Reconstructed Sea Surface Temperature (ERSST) version 4 dataset \citep{huang_extended_2015}.  NINO3.4 is defined as the average SST anomaly spanning 120–170$^\circ$W, 5$^\circ$S–5$^\circ$N.  All correlation coefficients are computed on detrended fields, with trends computed over 1979-2023 via linear regression and then subtracted off.  Regarding statistical significance, autocorrelations are small for all fields examined (not shown), justifying approximating each year of the 45 year record as independent.   The resulting threshold correlation coefficient magnitude for the \(p=0.05\) significance level is 0.29.

\section{Seasonal cycle}
\label{sec:seasonal}

\begin{figure}[h]
  \centering
  \includegraphics[width=0.5\textwidth]{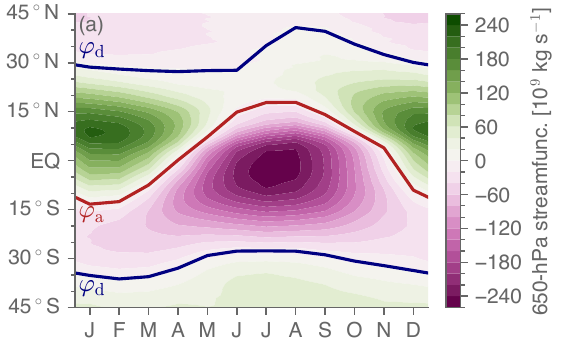}\\
  \includegraphics[width=0.5\textwidth]{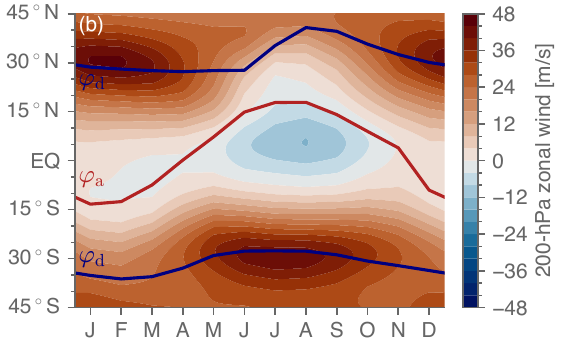}\\
  \includegraphics[width=0.5\textwidth]{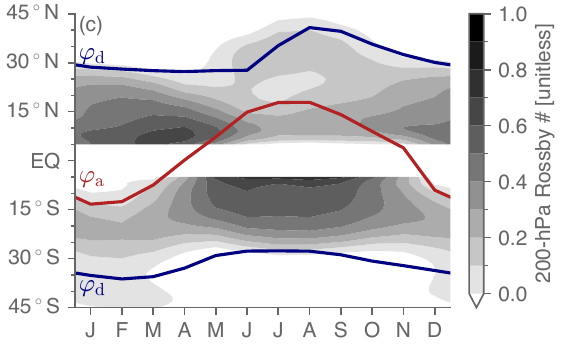}
  \caption{In solid curves in each panel, the Hadley cell ascending edge in dark red and the descending edges in dark blue.
    In color shading, climatological seasonal cycles of (a) meridional mass overturning streamfunction at 650 hPa, and at 200 hPa of (b) zonal wind and (c) the local Rossby number, with values in (c) masked out within 5\(^\circ\) of the equator where the denominator approaches zero.}
  \label{fig:sf-ro-u}
\end{figure}

Fig.~\ref{fig:sf-ro-u}a shows the climatological seasonal cycles of both descending edges and the ascending edge.  The SH descending edge ranges from 27.6$^\circ$S in July to 36.2$^\circ$S in February, and the NH descending edge ranges from 27.3$^\circ$N in April to 40.8$^\circ$N in August.  The ascending edge varies smoothly between 13.3\(^\circ\)S in January to 17.9\(^\circ\)N in August, lagging the insolation by $\sim$1 month owing to the coupled surface-lower atmosphere system's thermal inertia \citep[\eg/][]{mitchell_effects_2014}.

These seasonal migrations in the cell edges occur in concert with changes in the rate of mass overturned within the cells, shown as filled contours in Fig.~\ref{fig:sf-ro-u}a, with either cell's mass overturning strengthening from that hemisphere's autumn into winter as it grows meridionally and becomes cross-equatorial, and subsequently contracting and weakening in local spring through summer.  Peak mass overturning rates, exceeding \(240{\times}10^9\)~kg~s\(^{-1}\), occur at and just south of the equator in austral winter within the center of the cross-equatorial cell.  

Fig.~\ref{fig:sf-ro-u}b shows the climatological monthly zonal-mean zonal wind at 200~hPa.  It is easterly at the equator except in boreal winter \citep{zhang_seasonal_2022}.  Within the meridional extent of the Hadley cells, the easterlies peak in meridional extent and magnitude---respectively, \(\sim\)10\(^\circ\)S-20\(^\circ\)N and \(\sim\)12~m s\(^{-1}\)---in boreal summer within the strong, cross-equatorial cell \citep{bordoni_monsoons_2008}.  In every month, winds monotonically become more positive moving poleward from their deep-tropical minimum, reaching \(\sim\)40 m~s\(^{-1}\) at the descending edge of either winter, cross-equatorial cell.  The 200-hPa zonal wind is small in the vicinity of the ascending edge in all months, which coheres qualitatively with the assumption of vanishing zonal wind there in our theory.

Fig.~\ref{fig:sf-ro-u}c shows the climatological monthly-mean local Rossby number at 200 hPa, masked out within 5\(^\circ\) latitude of the equator where it is ill-defined.  It has a more complicated meridional structure than the zonal wind itself, though throughout the year in both hemispheres it is small, \(\sim\)0-0.2, in the vicinity of the descending edge.  It reaches up to \(\sim\)0.7 within either winter, cross-equatorial cell, supported by a reinforcing feedback: ascent moving poleward increases the extent and magnitude of upper-tropospheric easterlies, which in turn prevents extratropical-origin eddies from breaking over more of the tropics, enabling the Rossby number to grow and thus further strengthening the easterlies \citep{schneider_eddy-mediated_2008,bordoni_monsoons_2008}.  At the ascending edge, the seasonality is somewhat complicated, though local minima featuring the smallest Rossby number values, \(\sim\)0.1, coincide with its poleward-most extents in either summer hemisphere.  All together, the meridional variations in the Rossby number within the Hadley cells are smallest and non-monotonic in latitude within both cells within either summer hemisphere, whereas they decrease sharply and nearly monotonically moving poleward from the equator within either winter hemisphere.

Despite this meridional structure in the local Rossby number, treating it as uniform within each cell results in surprisingly accurate zonal wind fields.  Fig.~\ref{fig:fixed-ro-methods}d shows this for June, and for completeness Figs.~\ref{fig:app-uro-ann-seas} and \ref{fig:app-uro-mon} show results for the annual mean, all four seasons, and all twelve calendar months.  In all cases, within the span of each Hadley cell, the uniform-Rossby number zonal wind field corresponds fairly well with the actual zonal wind field.  Two limitations of the \(\uro\) field's accuracy are not capturing the leveling off of the zonal wind in the vicinity of the subtropical jets and, for November through February, being incapable of capturing the superrotation, \ie/ positive equatorial zonal wind \citep{zhang_seasonal_2022}, since \(\uro\) cannot exceed zero at the equator.

\begin{figure}
  \centering
  \includegraphics[width=0.5\textwidth]{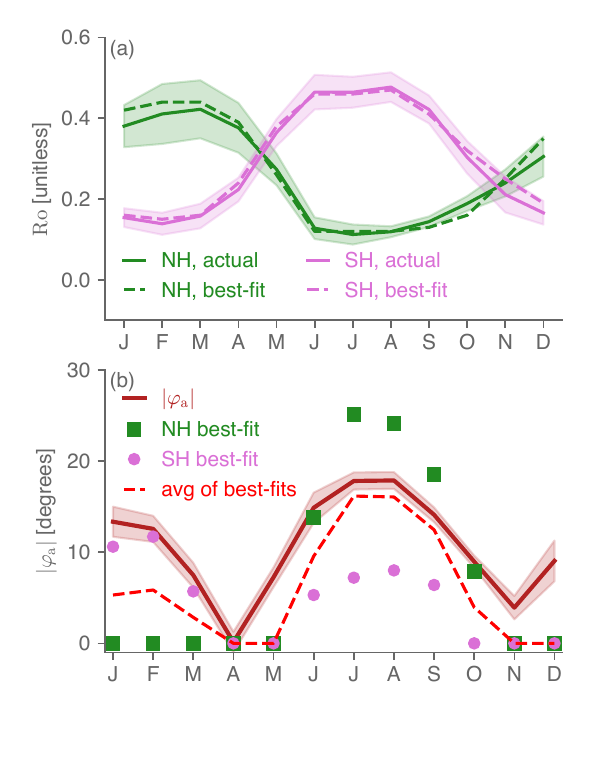}
  \caption{
    (a) Seasonal cycles of cell-mean $\Ro$.  Solid curves are the climatological values diagnosed directly from ERA5, with the \(\pm1\) standard deviation range shaded, and dashed curves are the values for the given month corresponding to the best-fit $\uro$ field against the ERA5 $u$.  (b) Seasonal cycle of $|\lata|$ in solid red, with the \(\pm1\) standard deviation range shaded.  Pink dots and green squares are the values corresponding to the SH and NH best-fit \(\uro\) fields, respectively, and the dashed red is their average.}
  \label{fig:uro-ro-lata-fits}
\end{figure}

Moreover, for each month, the Rossby number value generating this best fit to the zonal winds is close to the actual cell-mean Rossby number, as shown in Fig.~\ref{fig:uro-ro-lata-fits}a.  Both best-fit and actual cell-mean Rossby number values are strongly seasonal, ranging from \(\sim\)0.1 in either summer cell, to \(\sim\)0.2-0.3 in the equinoctial cells, and up to \(\sim\)0.4-0.5 in the cross-equatorial solsticial cells.
The increase in the Rossby number in either hemisphere from summer to winter reflects the transition from a regime in which eddy stresses prevail throughout the cell, reducing the Rossby number, to one where, as noted above, easterlies in the cross-equatorial cell shield low latitudes from these eddies, enabling the local Rossby number to increase \citep{schneider_eddy-mediated_2008,bordoni_monsoons_2008}.

We have also computed the maximum of the local Rossby number within each Hadley cell, over the same span of latitudes in each case as used to compute the cell-mean value.  The cell-maximum Rossby number reaches up to 0.65 for the NH in March and 0.79 for the SH in August, and for both cells it is very highly correlated with the cell-mean value: \(r=0.967\) for the NH and \(r=0.996\) for the SH (not shown).

Fig.~\ref{fig:uro-ro-lata-fits}b shows the monthly climatologies of the ascending edge poleward displacement, the corresponding values for each of the NH and SH from the best-fit \(\uro\) fields, and the average of these NH and SH values.  The best-fit ascending edge displacements broadly agree with the corresponding diagnosed values, albeit less cleanly than for the best-fit \vs/ cell-mean Rossby numbers.  The actual \(|\lata|\) expands poleward and contracts equatorward twice over the seasonal cycle, moving as far poleward as \({\sim}18^\circ\) in boreal summer and \({\sim}13^\circ\) in austral summer.  The best-fit values are closest to this in either summer hemisphere, and the average of the best-fit values in either hemisphere in each month is equatorward but otherwise comparable to the actual.  The NH best-fit ascending edge reaches as far as 25.1\(^\circ\)N during boreal summer, which likely results from the regional monsoons, especially the Indian summer monsoon, which enables ascent regionally to extend deeper into the subtropics \citep[e.g.][]{nie_observational_2010}.

We conclude from these diagnostics that a uniform-Rossby number approximation, though poor for the Rossby number itself, yields useful fields for our purposes: \(\uro\) captures the actual upper-branch zonal winds fairly well, with the corresponding \(\Ro\) and \(|\lata|\) values being consistent with the diagnosed ones.  We therefore proceed using the \(\uro\)-based theory for the descending edge based on baroclinic instability onset.

\begin{figure*}
  \centering
  \includegraphics[width=\textwidth]{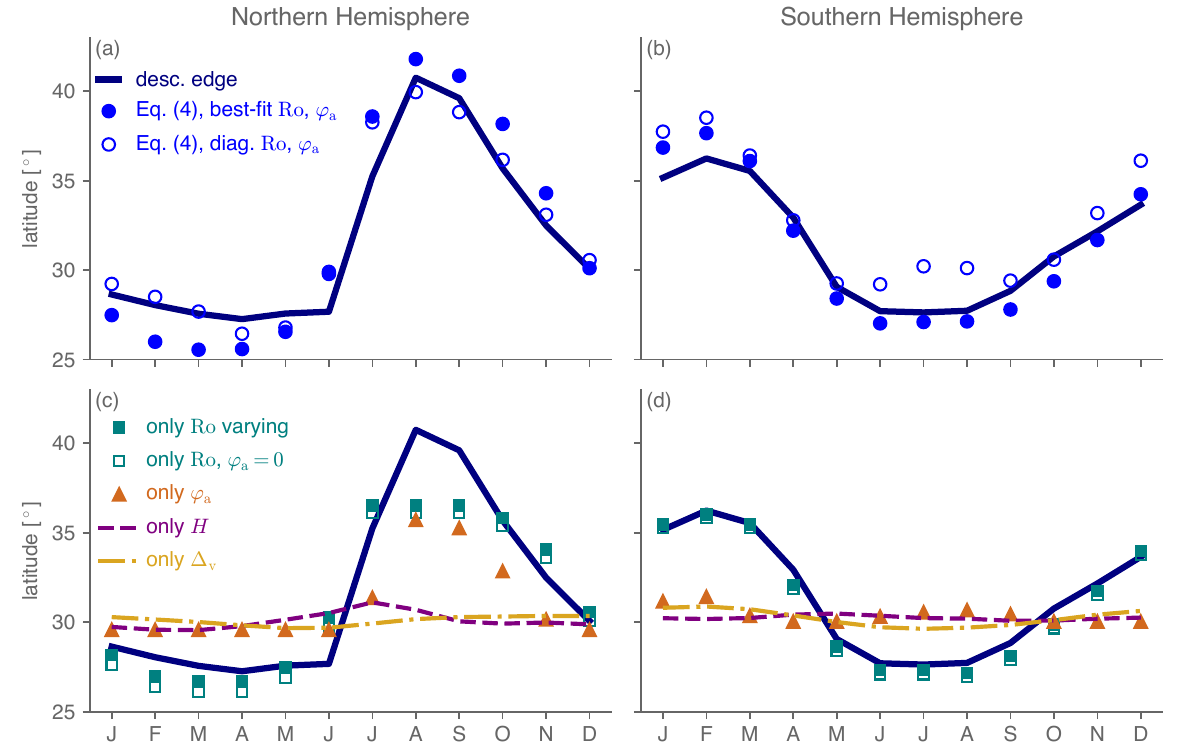}
\caption{Climatological monthly seasonal cycle of $\latd$ in ERA5 in the (a) NH and (b) SH. Solid navy is the actual field, and blue dots are our theoretical prediction using (\ref{eq:bci-edge-off-eq}), with the values of \(\Ro\) and \(\lata\) in each month coming from either (filled dots) the best-fit \(\uro\) fields or (unfilled) directly diagnosed.  In (c) and (d), additional curves use (\ref{eq:bci-edge-off-eq}) but with all terms except the specified one held constant to their average value across months.  For the NH, a lag of one month has been applied to all theoretical predictions.}
  \label{fig:latd-predic}
\end{figure*}

Fig.~\ref{fig:latd-predic}a,b show the seasonal cycles of the NH and SH descending edges and their predictions using (\ref{eq:bci-edge-off-eq}).  Two predictions are shown for each hemisphere, one using the best-fit values and one using the diagnosed values for \(\Ro\) and \(|\lata|\).  For the NH only, these include an empirical 1-month lag to the predictions, as \citet{hill_theory_2022} also required for both hemispheres in an idealized aquaplanet.  Our theory captures the seasonal cycle of the descending edge in each hemisphere well.  It performs slightly better using the best-fit rather than diagnosed \(\Ro\) and \(\lata\) values for the SH, but vice versa for the NH.  

Given this accuracy, we can evaluate the relative importance of each term by holding all terms in (\ref{eq:bci-edge-off-eq}) constant at their average value across months except one that remains fully varying, with the results shown in Fig.~\ref{fig:latd-predic}c,d.  These use the best-fit rather than diagnosed values for the Rossby number and ascending edge, but this makes little difference (not shown).  The seasonal cycles of the mid-latitude tropopause heights and static stabilities matter very little.  This is partly because their own seasonal cycles are both small, varying by at most a few percent from their averages across months (not shown).  Fractional variations in the ascending edge displacement are larger, yet its seasonality also plays a mostly modest role, with a substantial contribution only to the NH descending edge in late boreal summer through autumn when it helps to push the descending edge to its most poleward values.  It is the Rossby number seasonality that predominantly controls the seasonal cycles of the descending edges in both hemispheres, with the predictions weakly modified when the seasonality in all other terms is removed.  For the SH, if anything the fit improves slightly when the other terms are held fixed.


In addition to the ascending edge seasonality being unimportant, its annual-mean displacement is modest enough that setting \(\lata=0\) only weakly modifies the descending edge predictions (unfilled squares in Fig.~\ref{fig:latd-predic}).  In that case, (\ref{eq:bci-edge-off-eq}) reduces to (\ref{eq:bci-edge-eq}), from which it follows that the descending edges should vary as \(\Ro^{-1/4}\).  The exponents estimated from linear regression in log-log space between the diagnosed $|\latd|$ and cell-mean $\Ro$ are fairly close to this: again with the 1-month empirical lag applied for the NH only, the values for the NH and SH are -0.25 and -0.23 respectively using the best-fit \(\Ro\) values or -0.28 and -0.21 using the diagnosed cell-mean \(\Ro\) values.

For the NH cell during local summer to autumn, it may seem curious that, on the one hand, including \(\lata\) variations alone captures much of the behavior of the descending edge, yet on the other hand setting \(\lata=0\) makes little difference when only \(\Ro\) is allowed to vary.  We reconcile these results by examining the partial derivatives of (\ref{eq:bci-edge-off-eq}).  Ignoring the \(c_\mr{d}\) constant for simplicity, these are:
\begin{align*}
  \label{eq:latd-partial-derivs}
  \frac{\partial\latd^2}{\partial\Ro}&\approx-\frac{\lat_{\mr{d},0}^4}{2\Ro}\left(\frac{\lata^4}{4}+\lat_{\mr{d},0}^4\right)^{-1/2},\\
  \frac{\partial\latd^2}{\partial\lata}&\approx\lata\left[1+\frac{\lata^2}{2}\left(\frac{\lata^4}{4}+\lat_{\mr{d},0}^4\right)^{-1/2}\right].
\end{align*}
If we further treat \(2^{-1/2}|\lata|\ll\lat_{\mr{d},0}\), which though imperfect seems fair enough for these purposes, these reduce to
\begin{align*}
  \frac{\partial\latd^2}{\partial\Ro}&\approx-\frac{\lat_{\mr{d},0}^2}{2\Ro},\\
  \frac{\partial\latd^2}{\partial\lata}&\approx\lata,
\end{align*}
from which it is clear that \(\latd\) is considerably more sensitive to variations in \(\Ro\) than to variations in \(\lata\).

\section{Interannual variability}
\label{sec:interann}

We now show that the main results just presented for the climatological seasonal cycle also hold for interannual variability of both hemispheres' annual-mean cells: the Hadley cell upper-branch zonal wind fields are well captured by \(\uro\), the descending edge variations are well captured by (\ref{eq:bci-edge-off-eq}), and  within (\ref{eq:bci-edge-off-eq}) the cell-mean Rossby number is the most influential term.

\begin{figure}
  \centering
  \includegraphics[width=0.49\textwidth]{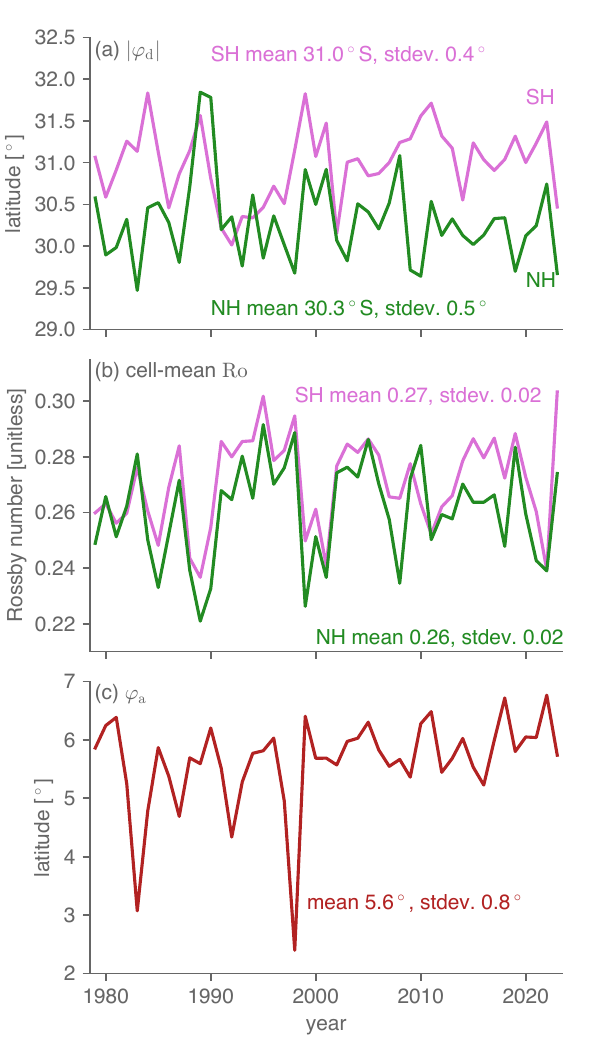}\\
  \caption{Timeseries spanning 1979-2023 of annual-mean values of (a) NH and SH Hadley cell descending edges, (b) NH and SH cell-mean Rossby numbers, and (c) the shared ascending edge.}
  \label{fig:interann-timeseries}
\end{figure}

Fig.~\ref{fig:interann-timeseries} shows the timeseries of the NH and SH descending edges, NH and SH cell-mean Rossby numbers, and the shared ascending edge, each computed from the annual-mean streamfunction for each calendar year, 1979-2023.  Both the mean and variance of each descending edge is similar between the hemispheres: means 30.3\(^\circ\)N and 31.0\(^\circ\)S and interannual standard deviations 0.49 for the NH and 0.43\(^\circ\) for the SH.  The covariation of the two descending edges is modest, with correlation coefficient \(r=0.30\). 
As for the seasonal cycle, the interannual edge displacements do not strongly depend on the metric used; comparing our definitions to the 500-hPa zero crossing, the correlation coefficients for the NH descending edge, ascending edge, and SH descending edge are 0.87, 0.95, and 0.96 respectively (not shown).

\begin{figure}
  \centering
  \includegraphics[width=0.49\textwidth]{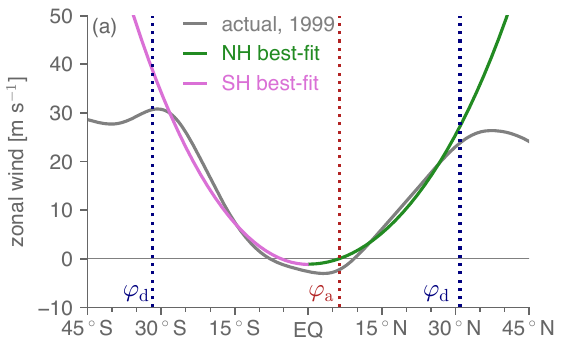}\\
  \includegraphics[width=0.49\textwidth]{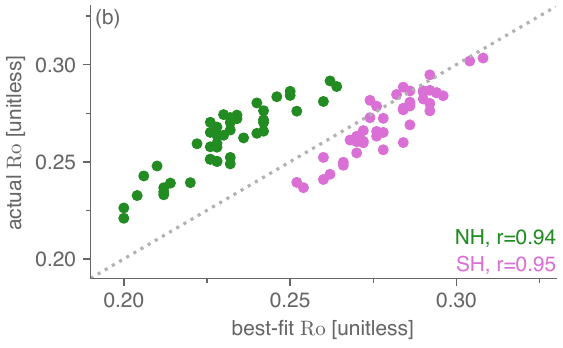}\\
  \includegraphics[width=0.49\textwidth]{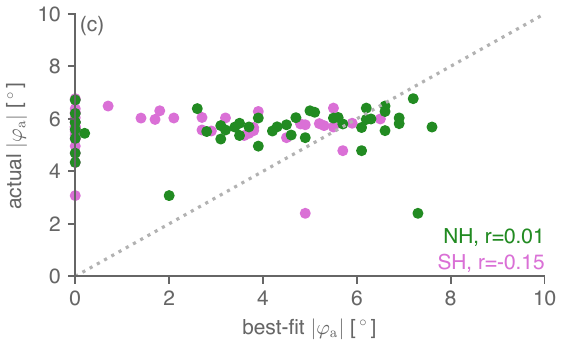}\\
  \caption{(a) For the year 1999, (grey) actual 200-hPa zonal-wind and corresponding best-fit \(\uro\) field for the (pink) SH and (green) NH Hadley cell, with the Hadley cell edges overlaid as vertical dotted lines.  (b) Scatterplot of best-fit vs. diagnosed cell-mean Rossby number in each year for the (pink) SH and (green) NH cells.  The corresponding Pearson's correlation coefficient for each hemisphere is also printed, and the dotted grey diagonal line is the one-to-one line.  (c) As in panel c, but for the best-fit \vs/ diagnosed ascending edge displacement rather than the Rossby number.}
  \label{fig:interann-scatter}
\end{figure}

For each year, we compute best-fit \(\uro\) profiles and corresponding best-fit Rossby number and ascending edge values for each Hadley cell for that year, following the same methodology as for the climatological seasonal cycle described above.  The uniform-Rossby approximations capture the upper-tropospheric zonal winds within the Hadley cells reasonably well, which Fig.~\ref{fig:interann-scatter}a shows for the year 1999, selected quasi-randomly, as one example and with the fits comparably accurate in all other years (not shown).  Fig.~\ref{fig:interann-scatter}b shows the best-fit and actual cell-mean Rossby numbers for each year plotted against one another; they are very highly correlated, \({r=0.94}\) and 0.95 for the NH and SH respectively, and fall close to the one-to-one line.  In contrast and unlike for the seasonal cycle, there is almost no correspondence between the actual and best-fit values for the ascending edge displacement, as shown in Fig.~\ref{fig:interann-scatter}c.  This implies that interannual variations in the annual-mean zonal wind in the Hadley cell upper branches depend weakly on interannual variations of the ascending edge (Fig.~\ref{fig:interann-timeseries}c).  This is consistent in sign with the annual-mean \(\lata\) itself being small, given that from (\ref{eq:uro}) we have \(\partial\uro/\partial\lata=-2\Ro\Omega a\lata\), \ie/ that the change in \(\uro\) with respect to \(\lata\) is proportional to \(\lata\) itself.

\begin{figure*}
  \centering
  \includegraphics[width=\textwidth]{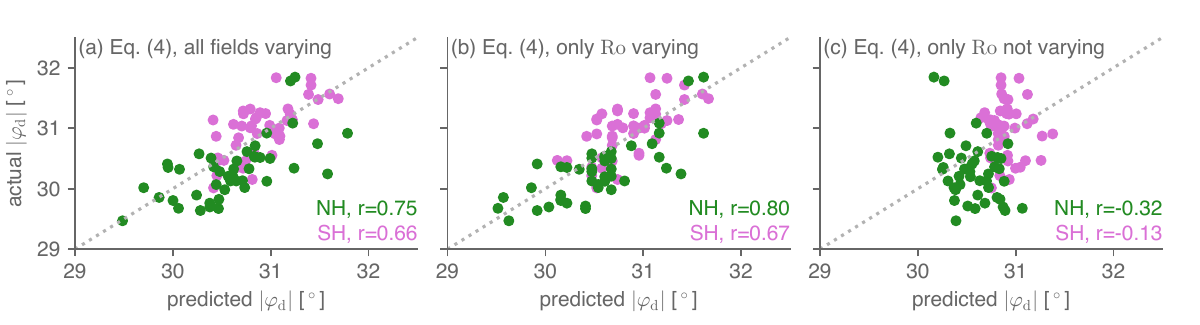}
  \caption{(a) Descending edge predicted by (\ref{eq:bci-edge-off-eq}) for each year using diagnosed values of all terms for that year and cell vs. the actual descending edge for that year and cell.  (b) As in (a), but with all fields except \(\Ro\) set to their climatological values.  (c) As in (a), but with only \(\Ro\) set to its climatological value.}
  \label{fig:interann-ro-sens}
\end{figure*}

We then use (\ref{eq:bci-edge-off-eq}) to predict the descending edge of each cell in each year using these best-fit values of \(\Ro\) and \(|\lata|\) and diagnosed values of the mid-latitude tropopause heights and static stabilities, with as noted \(c_\mr{d}\) values of 1.02 for the NH and 1.09 for the SH.  Fig.~\ref{fig:interann-ro-sens}a shows the results.  Our theory performs fairly well, \({r=0.75}\) and 0.66 for the NH and SH, respectively, though it clearly leaves more variability unexplained than in the seasonal cycle case.  Unlike the seasonal cycle, the theory performs somewhat better overall for the NH compared to the SH.  As expected for annual averages, no empirical lag is required for either hemisphere.  Results are similar using the diagnosed rather than best-fit \(\Ro\) and \(|\lata|\) values: \({r=0.77}\) and 0.57 between the prediction and the actual edge for the NH and SH, respectively.

As for the seasonal cycle, the variations in \(\Ro\) predominate over changes in all other fields in setting the descending edge.  This can be seen in Fig.~\ref{fig:interann-ro-sens}b and c, which show the predicted \vs/ actual descending edges in each year when either (panel b) \(\Ro\) is the only term varying or (panel c) \(\Ro\) is the only term \emph{not} varying.  In fact the fits are slightly improved when the factors other than the Rossby number are held constant (\({r=0.80}\) and 0.67 for NH and SH, respectively), as the relationship between the descending edges and all the other factors combined, though weak, are of the wrong sign for both hemispheres  (\({r=-0.32}\) and -0.13 for NH and SH, respectively).  This stems from the tropopause height: the descending edges are negatively correlated with the mid-latitude tropopause height in the same hemisphere (not shown).  However, this is sensitive to the definition of the mid-latitude band; when fixed boundaries are used instead, \eg/ 40-60\(^\circ\), the correlations become insignificant.  In any case, by any metric the mid-latitude tropopause height is not an important factor.

The influence of each factor can be understood as follows.  The mid-latitude tropopause heights and static stabilities simply do not change much from year to year: none varies by more than \(\pm\)3\% for any year 1979-2023 from its climatological value (not shown).
Over the same period, the cell-mean Rossby numbers vary by as much as \(\pm\)15\%.
For the ascending edge displacement, though fractional variations from year to year can exceed \(\sim\)50\%, in absolute terms these displacements are smaller than those during the seasonal cycle---minimum and maximum values across all years are 2.4 and 6.8\(^\circ\), respectively, compared to corresponding seasonal cycle values of 0.1 and 17.7\(^\circ\)---in which context also they play a minor role compared to Rossby number variations.  

This predominance of the Rossby number means that, again as for the seasonal cycle, the full expression (\ref{eq:bci-edge-off-eq}) can be effectively replaced with the on-equatorial-ascent case (\ref{eq:bci-edge-eq}), yielding an expected power-law exponent of -1/4 for \(\Ro\) against each descending edge.  The power-law exponents estimated from the log-log regression of \(\latd\) against \(\Ro\) are somewhat lower than, but in the ballpark of, this -0.25 value: -0.19 for the NH and -0.21 for the SH using the best-fit \(\Ro\) or -0.19 and -0.15 using the diagnosed cell-mean \(\Ro\).

\begin{figure}
  \centering
  \includegraphics[width=0.5\textwidth]{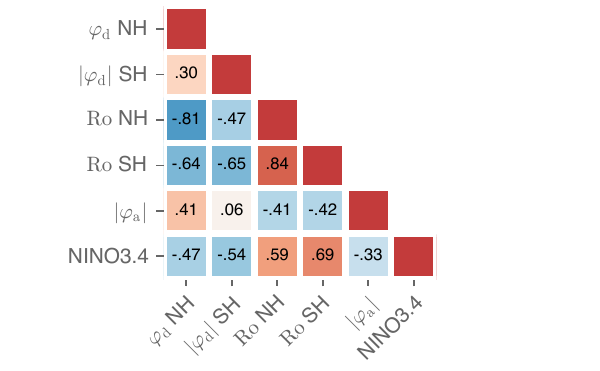}
  \caption{Heat map of Pearson correlation coefficients among various metrics of the NH and SH Hadley cells as well as the NINO3.4 index, all annual means.  From top to bottom and left to right in each panel, the quantities are: NH cell-mean $\Ro$, SH cell-mean $\Ro$, NH $\latd$, $|\lata|$, SH $-\latd$, and the NINO3.4 index.  All quantities are computed over 1979-2023 and linearly detrended over that period prior to the computation of correlations.}
  \label{fig:interann-heatmap}
\end{figure}

The Rossby numbers for the NH and SH cells (Fig.~\ref{fig:interann-timeseries}b) are highly correlated with one another (${r=0.84}$), indicating that interannual variations in the eddy stresses tend to be coherent across the Tropics, evocative of the tropics-wide influence of ENSO \citep{lu_response_2008}.  Fig.~\ref{fig:interann-heatmap} provides an interannual correlation heat map of the diagnosed cell-mean Rossby numbers, descending edges, ascending edge displacement, and the standard NINO3.4 index, with all correlations computed using linearly detrended fields.  Consistent in sign with our arguments and the existing literature regarding ENSO, \elnino/ conditions (\ie/ positive NINO3.4) act to contract all three Hadley cell edges equatorward and increase both cell-mean Rossby numbers.  Conversely, the SH descending edge has virtually no linear relationship with the ascending edge (\({r=0.06}\)).  While prior literature has argued that the ENSO relationships with the Hadley cells are tighter for the SH than for the NH \citep{hasan_hemisphere-dependent_2024}, with the Arctic Oscillation playing a comparable role to ENSO for the NH \citep{seo_what_2023}, the difference between the hemispheres we find in this regard is modest, with \({r=-0.47}\) \vs/ -0.54 between NINO3.4 and the descending edge for the NH and SH, respectively.

Both cell-mean Rossby numbers decrease as the ascending edge moves farther off equator (\({r=-0.41}\) and -0.42 for the NH and SH respectively), whereas for the seasonal cycle the poleward migration of \(|\lata|\) acts to increase \(\Ro\) within the cross-equatorial cell by generating stronger and more extensive easterlies that shield the deep tropics from extratropical wave breaking \citep{schneider_eddy-mediated_2008}.  In the annual mean, the SH cell is modestly cross-equatorial, and so by this mechanism alone the SH cell-mean Rossby number would be positively rather than negatively correlated with the ascending edge displacement.  Arguably, this mechanism becomes salient only once the cell becomes even more monsoonal, and a more compelling explanation lies in ENSO: in El Ni\~no years, the anomalous surface warmth in the eastern equatorial Pacific directly pulls the ascent equatorward \citep[\eg/][]{adam_seasonal_2016} and also, through mechanisms we are less certain of, enables the Rossby number within the Tropics to grow.

\section{Conclusions}
\label{sec:conc}

We conclude that both the climatological seasonal cycle and interannual variations of Earth's poleward, descending Hadley cell edges can be usefully interpreted via the simple two-layer, quasi-geostrophic baroclinic instability expression (\ref{eq:bci-edge-off-eq}), in which crucially the upper-branch zonal wind profile used within either cell is the uniform-Rossby number solution (\ref{eq:uro}).  The value of this uniform Rossby number amounts to a cell-mean value of the local Rossby number across each cell's upper branch.  And of the terms appearing in (\ref{eq:bci-edge-off-eq}), this cell-mean Rossby number predominantly controls seasonal and interannual variations of both hemispheres' descending edges.

With these results from ERA5, the uniform-\(\Ro\) framework for the descending edges encapsulated into (\ref{eq:bci-edge-off-eq}) has now proven successful for two distinct regimes of Hadley cell seasonality.  \citet{hill_theory_2022} originally test the theory using a seasonally forced aquaplanet simulation with a rather shallow, 10-m ocean mixed-layer depth.  As a result, the Hadley cells---though Earth-like in their overall meridional extent---are Mars-like in their seasonality \citep{zalucha_analysis_2010}: \(\lata\) migrates abruptly and deeply into either summer hemisphere just after either equinox, so much so that \(\lata\) rather than \(\Ro\) predominantly controls the descending edge migrations.  The theory's dexterity across these regimes suggests it could be useful for the Hadley cells of terrestrial atmospheres broadly.   

At the same time, in addition to lacking a predictive theory for \(\Ro\)---which, for Earth at least, is the central parameter---we do not understand why the theory's predictions lead the actual descending edge by one month in the NH only.  It can be inferred from \citet[][\cf/ their Fig.~6]{peles_estimating_2023} that multiple other metrics of the lowest latitude of baroclinic growth also lead the Hadley cell descending edge by roughly one month in the NH but not the SH.  This suggests roots in something other than the precise physical assumptions we've made.  The same lag is required in the \cite{hill_theory_2022} aquaplanet case just described---meaning, somewhat curiously, it is the case with the largest heat capacity, Earth's SH, in which the descending edge is more in phase with these environmental influences.  \citet{hill_theory_2022} also speculate about the finite timescale of Rossby wave generation in the extratropics followed by equatorward propagation and breaking; for that mechanism to be relevant here, wave generation and propagation would need to occur more slowly in the NH than SH.  Analysis of idealized simulations with differing mixed-layer depths between the two hemispheres or idealized continents \citep[\eg/][]{maroon_precipitation_2016,voigt_tropical_2016,hui_response_2021} would help shed light on all this.


Notably uninfluential both seasonally and interannually is the mid-latitude static stability, which under global warming is projected to increase and, through appeal to (\ref{eq:bci-edge-eq}), thereby predominantly drive the poleward expansion of the descending edges \citep[\eg/][]{kang_expansion_2012,chemke_exploiting_2019}.  To our knowledge, no study has yet investigated the role of the cell-mean Rossby numbers in warming-driven Hadley cell expansion; \citet{kang_expansion_2012} find large seasonal differences in the climatological value of a bulk Rossby number for each Hadley cell, but they neglect any changes to these bulk Rossby numbers as the planet warms.  All else equal, an increase in the Rossby number would retract the descending edge, while a decrease would expand the descending edge poleward.  The influence of the Rossby number on Hadley cell expansion with global warming could also come from the Rossby number's climatological value rather than its own response to warming: by (\ref{eq:bci-edge-eq}), an atmosphere with larger climatological cell-mean Rossby numbers would migrate less for a given change in any of the other terms, compared to a model with smaller climatological cell-mean Rossby numbers.  Given that multi-model-mean projections under high-CO2 forcing are robustly of expansion of the descending edges \citep[\eg/][]{grise_hadley_2020}, possibilities include that the cell-mean Rossby numbers are increasing, that they are decreasing but not enough to counter the increased subtropical static stability, or that the framework simply doesn't work in this context.


\clearpage
\acknowledgments
We thank Isaac Held for useful discussions of this work.  S.A.H. acknowledges funding from the NSF Climate and Large-scale Dynamics program, award AGS-2411723.  J.L.M. acknowledges funding from the NSF Climate and Large-scale Dynamics program, award AGS-1912673.  S.B. acknowledges partial support from the European Union under Next Generation EU, Mission 4 Component 2 - CUP E53D23021930001 and Next Generation EU, Mission 4 Component 2 - CUP E63C22000970007.

\datastatement
ERA5 data is available for free, public download at \url{https://www.ecmwf.int/en/forecasts/dataset/ecmwf-reanalysis-v5}.

\begin{figure*}
  \center
  \includegraphics[width=\textwidth]{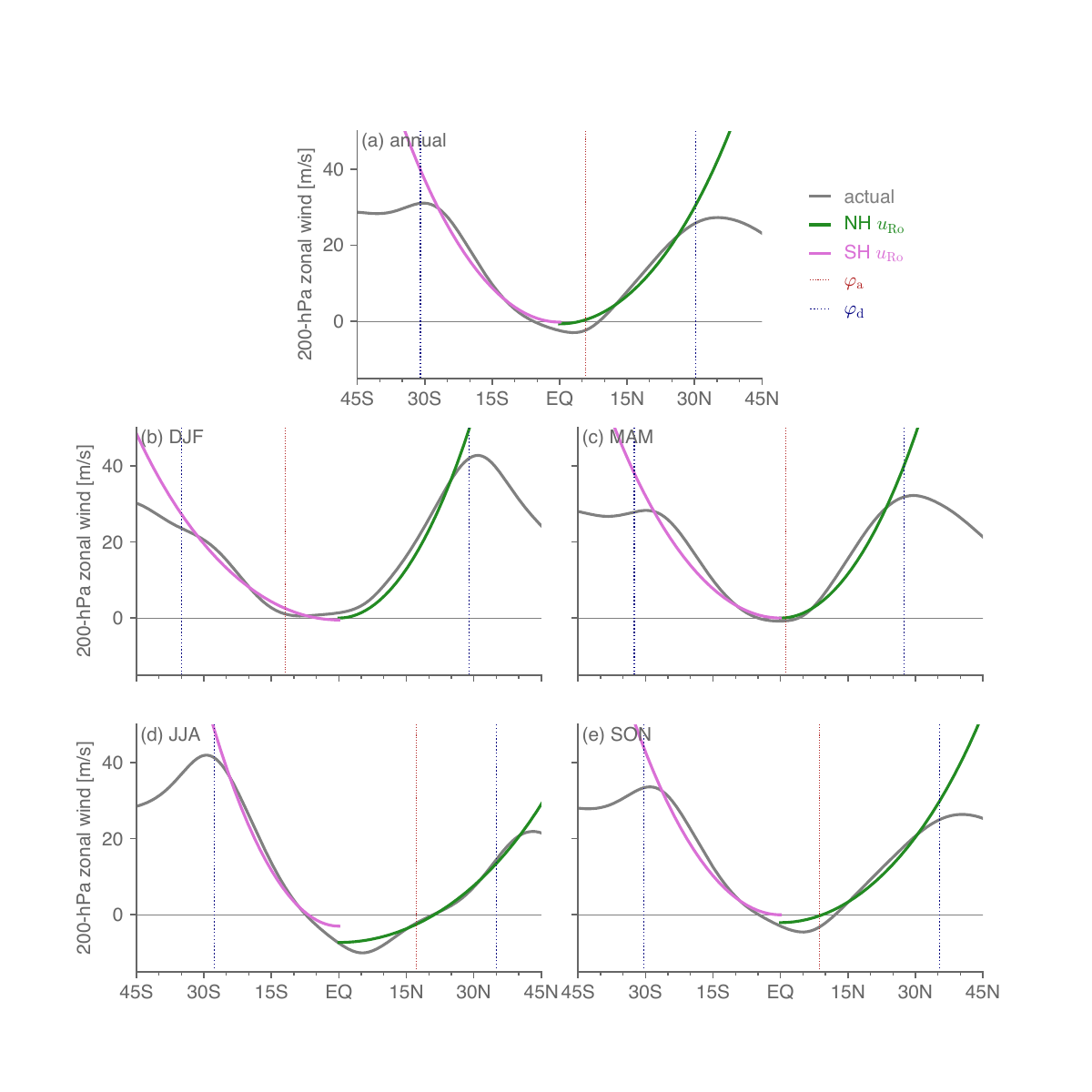}
  \caption{For the (a) annual and (b-e) seasonal climatologies, (grey) 200-hPa zonal wind and best-fit uniform-\(\Ro\) zonal wind fields for the (pink) SH and (green) NH Hadley cell, restricting to that hemisphere.  Overlaid red and blue dotted vertical lines are the diagnosed ascending edge and descending edges, respectively.}
  \label{fig:app-uro-ann-seas}
\end{figure*}

\begin{figure*}
  \center
  \includegraphics[width=\textwidth]{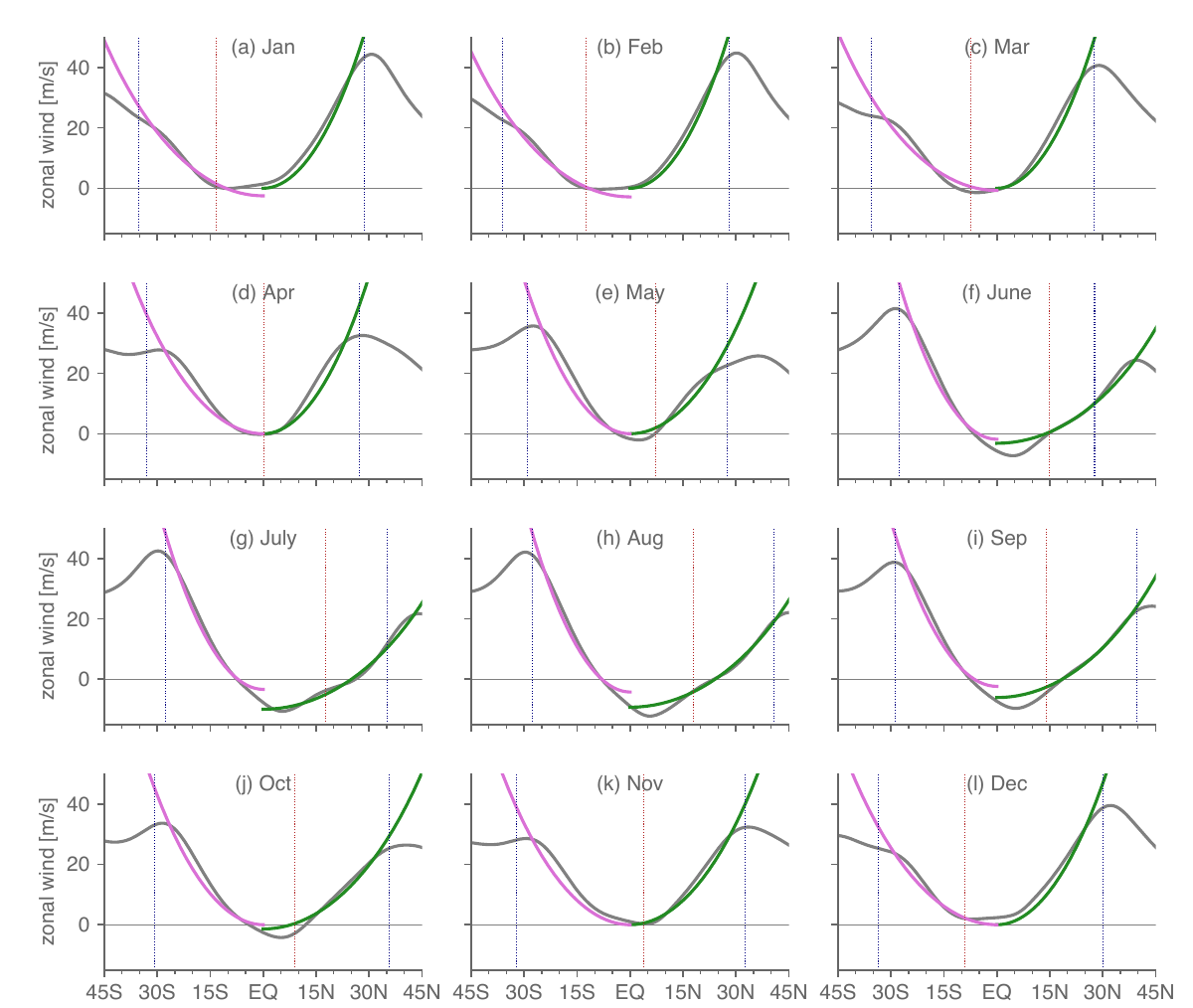}
  \caption{Same as Fig.~\ref{fig:app-uro-ann-seas}, but for each calendar month.}
  \label{fig:app-uro-mon}
\end{figure*}

\appendix
\appendixtitle{Uniform-Rossby number zonal wind fields compared to actual fields}
For completeness, Fig.~\ref{fig:app-uro-ann-seas} shows the diagnosed ERA5 200-hPa zonal wind and corresponding best-fit \(\uro\) fields computed for the climatological annual mean and for each meteorological season, and Fig.~\ref{fig:app-uro-mon} shows the same for each calendar month.

\clearpage

\clearpage

\clearpage
\bibliographystyle{ametsocV6}
\bibliography{./references}

\begin{thebibliography}{53}
\providecommand{\natexlab}[1]{#1}
\providecommand{\url}[1]{\texttt{#1}}
\renewcommand{\UrlFont}{\rmfamily}
\providecommand{\urlprefix}{URL }
\expandafter\ifx\csname urlstyle\endcsname\relax
  \providecommand{\doi}[1]{https://doi.org/\discretionary{}{}{}#1}\else
  \providecommand{\doi}{https://doi.org/\discretionary{}{}{}\begingroup
  \urlstyle{rm}\Url}\fi
\providecommand{\eprint}[2][]{\url{#2}}

\bibitem[{Adam et~al.(2016)Adam, Bischoff,, and Schneider}]{adam_seasonal_2016}
Adam, O., T.~Bischoff, and T.~Schneider, 2016: Seasonal and {{Interannual
  Variations}} of the {{Energy Flux Equator}} and {{ITCZ}}. {{Part I}}:
  {{Zonally Averaged ITCZ Position}}. \textit{Journal of Climate},
  \textbf{29~(9)}, 3219--3230, \doi{10.1175/JCLI-D-15-0512.1}.

\bibitem[{Adam et~al.(2018)}]{adam_tropd_2018}
Adam, O., and Coauthors, 2018: The {{TropD}} software package (v1):
  Standardized methods for calculating tropical-width diagnostics.
  \textit{Geoscientific Model Development}, \textbf{11~(10)}, 4339--4357,
  \doi{10.5194/gmd-11-4339-2018}.

\bibitem[{Bordoni and Schneider(2008)Bordoni, and
  Schneider}]{bordoni_monsoons_2008}
Bordoni, S., and T.~Schneider, 2008: Monsoons as eddy-mediated regime
  transitions of the tropical overturning circulation. \textit{Nature
  Geoscience}, \textbf{1~(8)}, 515--519, \doi{10.1038/ngeo248}.

\bibitem[{Caballero(2007)}]{caballero_role_2007}
Caballero, R., 2007: Role of eddies in the interannual variability of
  {{Hadley}} cell strength. \textit{Geophysical Research Letters},
  \textbf{34~(22)}, L22\,705, \doi{10.1029/2007GL030971}.

\bibitem[{Chemke and Polvani(2019)Chemke, and Polvani}]{chemke_exploiting_2019}
Chemke, R., and L.~M. Polvani, 2019: Exploiting the {{Abrupt}}
  4{\texttimes}{{CO2 Scenario}} to {{Elucidate Tropical Expansion Mechanisms}}.
  \textit{Journal of Climate}, \textbf{32~(3)}, 859--875,
  \doi{10.1175/JCLI-D-18-0330.1}.

\bibitem[{Chen and Held(2007)Chen, and Held}]{chen_phase_2007}
Chen, G., and I.~M. Held, 2007: Phase speed spectra and the recent poleward
  shift of {{Southern Hemisphere}} surface westerlies. \textit{Geophysical
  Research Letters}, \textbf{34~(21)}, \doi{10.1029/2007GL031200}.

\bibitem[{Davis and Birner(2022)Davis, and Birner}]{davis_eddy-mediated_2022}
Davis, N.~A., and T.~Birner, 2022: Eddy-{{Mediated Hadley Cell Expansion}} due
  to {{Axisymmetric Angular Momentum Adjustment}} to {{Greenhouse Gas
  Forcings}}. \textit{Journal of the Atmospheric Sciences}, \textbf{79~(1)},
  141--159, \doi{10.1175/JAS-D-20-0149.1}.

\bibitem[{Dima and Wallace(2003)Dima, and Wallace}]{dima_seasonality_2003}
Dima, I.~M., and J.~M. Wallace, 2003: On the {{Seasonality}} of the {{Hadley
  Cell}}. \textit{Journal of the Atmospheric Sciences}, \textbf{60~(12)},
  1522--1527, \doi{10.1175/1520-0469(2003)060<1522:OTSOTH>2.0.CO;2}.

\bibitem[{Ferrel(1856)}]{ferrel_essay_1856}
Ferrel, W., 1856: An essay on the winds and the currents of the ocean.
  \textit{Nashville Journal of Medicine and Surgery}, \textbf{11}, 13.

\bibitem[{Geen et~al.(2019)Geen, Lambert,, and Vallis}]{geen_processes_2019}
Geen, R., F.~H. Lambert, and G.~K. Vallis, 2019: Processes and {{Timescales}}
  in {{Onset}} and {{Withdrawal}} of ``{{Aquaplanet Monsoons}}''.
  \textit{Journal of the Atmospheric Sciences}, \textbf{76~(8)}, 2357--2373,
  \doi{10.1175/JAS-D-18-0214.1}.

\bibitem[{Grise and Davis(2020)Grise, and Davis}]{grise_hadley_2020}
Grise, K.~M., and S.~M. Davis, 2020: Hadley cell expansion in {{CMIP6}} models.
  \textit{Atmospheric Chemistry and Physics}, \textbf{20~(9)}, 5249--5268,
  \doi{10.5194/acp-20-5249-2020}.

\bibitem[{Hadley(1735)}]{hadley_concerning_1735}
Hadley, {\relax Geo}., 1735: Concerning the {{Cause}} of the {{General
  Trade-Winds}}. \textit{Philosophical Transactions of the Royal Society of
  London}, \textbf{39~(436-444)}, 58--62, \doi{10.1098/rstl.1735.0014}.

\bibitem[{Halley(1686)}]{halley_historical_1686}
Halley, E., 1686: An {{Historical Account}} of the {{Trade Winds}}, and
  {{Monsoons}}, {{Observable}} in the {{Seas}} between and {{Near}} the
  {{Tropicks}}, with an {{Attempt}} to {{Assign}} the {{Phisical Cause}} of the
  {{Said Winds}}. \textit{Philosophical Transactions of the Royal Society of
  London}, \textbf{16~(179-191)}, 153--168, \doi{10.1098/rstl.1686.0026}.

\bibitem[{Hasan et~al.(2024)Hasan, Larson, McMonigal, Robinson,, and
  Aiyyer}]{hasan_hemisphere-dependent_2024}
Hasan, M., S.~M. Larson, K.~McMonigal, W.~A. Robinson, and A.~Aiyyer, 2024:
  Hemisphere-{{Dependent Impacts}} of {{ENSO}} and {{Atmospheric Eddies}} on
  {{Hadley Circulation}}. \textit{Journal of Climate}, \textbf{37~(24)},
  6533--6548, \doi{10.1175/JCLI-D-24-0112.1}.

\bibitem[{Held(1978)}]{held_vertical_1978}
Held, I.~M., 1978: The {{Vertical Scale}} of an {{Unstable Baroclinic Wave}}
  and {{Its Importance}} for {{Eddy Heat Flux Parameterizations}}.
  \textit{Journal of the Atmospheric Sciences}, \textbf{35~(4)}, 572--576,
  \doi{10.1175/1520-0469(1978)035<0572:TVSOAU>2.0.CO;2}.

\bibitem[{Held(2000)}]{held_general_2000}
Held, I.~M., 2000: The {{General Circulation}} of the {{Atmosphere}}.
  \textit{The {{General Circulation}} of the {{Atmosphere}}: 2000 {{Program}}
  in {{Geophysical Fluid Dynamics}}}, No. WHOI-2001-03, Woods {{Hole Oceanog}}.
  {{Inst}}. {{Tech}}. {{Rept}}., Woods Hole Oceanographic Institution, 1--54.

\bibitem[{Held and Hou(1980)Held, and Hou}]{held_nonlinear_1980}
Held, I.~M., and A.~Y. Hou, 1980: Nonlinear axially symmetric circulations in a
  nearly inviscid atmosphere. \textit{Journal of the Atmospheric Sciences},
  \textbf{37~(3)}, 515--533,
  \doi{10.1175/1520-0469(1980)037<0515:NASCIA>2.0.CO;2}.

\bibitem[{Hersbach et~al.(2020)}]{hersbach_era5_2020}
Hersbach, H., and Coauthors, 2020: The {{ERA5}} global reanalysis.
  \textit{Quarterly Journal of the Royal Meteorological Society},
  \textbf{146~(730)}, 1999--2049, \doi{10.1002/qj.3803}.

\bibitem[{Hilgenbrink and Hartmann(2018)Hilgenbrink, and
  Hartmann}]{hilgenbrink_response_2018}
Hilgenbrink, C.~C., and D.~L. Hartmann, 2018: The {{Response}} of {{Hadley
  Circulation Extent}} to an {{Idealized Representation}} of {{Poleward Ocean
  Heat Transport}} in an {{Aquaplanet GCM}}. \textit{Journal of Climate},
  \textbf{31~(23)}, 9753--9770, \doi{10.1175/JCLI-D-18-0324.1}.

\bibitem[{Hill et~al.(2021)Hill, Bordoni,, and Mitchell}]{hill_solsticial_2021}
Hill, S.~A., S.~Bordoni, and J.~L. Mitchell, 2021: Solsticial {{Hadley Cell
  Ascending Edge Theory}} from {{Supercriticality}}. \textit{Journal of the
  Atmospheric Sciences}, \textbf{78~(6)}, 1999--2011,
  \doi{10.1175/JAS-D-20-0341.1}.

\bibitem[{Hill et~al.(2022)Hill, Bordoni,, and Mitchell}]{hill_theory_2022}
Hill, S.~A., S.~Bordoni, and J.~L. Mitchell, 2022: A {{Theory}} for the
  {{Hadley Cell Descending}} and {{Ascending Edges}} throughout the {{Annual
  Cycle}}. \textit{Journal of the Atmospheric Sciences}, \textbf{79~(10)},
  2515--2528, \doi{10.1175/JAS-D-21-0328.1}.

\bibitem[{Huang et~al.(2015)}]{huang_extended_2015}
Huang, B., and Coauthors, 2015: Extended {{Reconstructed Sea Surface
  Temperature Version}} 4 ({{ERSST}}.v4). {{Part I}}: {{Upgrades}} and
  {{Intercomparisons}}. \textit{Journal of Climate}, \textbf{28~(3)}, 911--930,
  \doi{10.1175/JCLI-D-14-00006.1}.

\bibitem[{Hui and Bordoni(2021)Hui, and Bordoni}]{hui_response_2021}
Hui, K.~L., and S.~Bordoni, 2021: Response of {{Monsoon Rainfall}} to
  {{Changes}} in the {{Latitude}} of the {{Equatorward Coastline}} of a
  {{Zonally Symmetric Continent}}. \textit{Journal of the Atmospheric
  Sciences}, \textbf{78~(5)}, 1429--1444, \doi{10.1175/JAS-D-20-0110.1}.

\bibitem[{Kang and Lu(2012)Kang, and Lu}]{kang_expansion_2012}
Kang, S.~M., and J.~Lu, 2012: Expansion of the {{Hadley Cell}} under global
  warming: {{Winter}} versus summer. \textit{Journal of Climate},
  \textbf{25~(24)}, 8387--8393, \doi{10.1175/JCLI-D-12-00323.1}.

\bibitem[{Korty and Schneider(2008)Korty, and Schneider}]{korty_extent_2008}
Korty, R.~L., and T.~Schneider, 2008: Extent of {{Hadley}} circulations in dry
  atmospheres. \textit{Geophysical Research Letters}, \textbf{35~(23)},
  L23\,803, \doi{10.1029/2008GL035847}.

\bibitem[{Levine and Schneider(2011)Levine, and
  Schneider}]{levine_response_2011}
Levine, X.~J., and T.~Schneider, 2011: Response of the {{Hadley Circulation}}
  to {{Climate Change}} in an {{Aquaplanet GCM Coupled}} to a {{Simple
  Representation}} of {{Ocean Heat Transport}}. \textit{Journal of the
  Atmospheric Sciences}, \textbf{68~(4)}, 769--783,
  \doi{10.1175/2010JAS3553.1}.

\bibitem[{Levine and Schneider(2015)Levine, and
  Schneider}]{levine_baroclinic_2015}
Levine, X.~J., and T.~Schneider, 2015: Baroclinic {{Eddies}} and the {{Extent}}
  of the {{Hadley Circulation}}: {{An Idealized GCM Study}}. \textit{Journal of
  the Atmospheric Sciences}, \textbf{72~(7)}, 2744--2761,
  \doi{10.1175/JAS-D-14-0152.1}.

\bibitem[{Lindzen and Hou(1988)Lindzen, and Hou}]{lindzen_hadley_1988}
Lindzen, R.~S., and A.~V. Hou, 1988: Hadley circulations for zonally averaged
  heating centered off the equator. \textit{Journal of the Atmospheric
  Sciences}, \textbf{45~(17)}, 2416--2427,
  \doi{10.1175/1520-0469(1988)045<2416:HCFZAH>2.0.CO;2}.

\bibitem[{Lu et~al.(2008)Lu, Chen,, and Frierson}]{lu_response_2008}
Lu, J., G.~Chen, and D.~M.~W. Frierson, 2008: Response of the {{Zonal Mean
  Atmospheric Circulation}} to {{El Ni{\~n}o}} versus {{Global Warming}}.
  \textit{Journal of Climate}, \textbf{21~(22)}, 5835--5851,
  \doi{10.1175/2008JCLI2200.1}.

\bibitem[{Maroon et~al.(2016)Maroon, Frierson, Kang,, and
  Scheff}]{maroon_precipitation_2016}
Maroon, E.~A., D.~M.~W. Frierson, S.~M. Kang, and J.~Scheff, 2016: The
  {{Precipitation Response}} to an {{Idealized Subtropical Continent}}.
  \textit{Journal of Climate}, \textbf{29~(12)}, 4543--4564,
  \doi{10.1175/JCLI-D-15-0616.1}.

\bibitem[{Mitchell et~al.(2014)Mitchell, Vallis,, and
  Potter}]{mitchell_effects_2014}
Mitchell, J.~L., G.~K. Vallis, and S.~F. Potter, 2014: Effects of the
  {{Seasonal Cycle}} on {{Superrotation}} in {{Planetary Atmospheres}}.
  \textit{The Astrophysical Journal}, \textbf{787~(1)}, 23,
  \doi{10.1088/0004-637X/787/1/23}.

\bibitem[{Nie et~al.(2010)Nie, Boos,, and Kuang}]{nie_observational_2010}
Nie, J., W.~R. Boos, and Z.~Kuang, 2010: Observational {{Evaluation}} of a
  {{Convective Quasi-Equilibrium View}} of {{Monsoons}}. \textit{Journal of
  Climate}, \textbf{23~(16)}, 4416--4428, \doi{10.1175/2010JCLI3505.1}.

\bibitem[{Peles and Lachmy(2023)Peles, and Lachmy}]{peles_estimating_2023}
Peles, O., and O.~Lachmy, 2023: Estimating the {{Lowest Latitude}} of
  {{Baroclinic Growth}}. \textit{Journal of the Atmospheric Sciences},
  \textbf{80~(5)}, 1401--1414, \doi{10.1175/JAS-D-22-0201.1}.

\bibitem[{Schneider(2006)}]{schneider_general_2006}
Schneider, T., 2006: The general circulation of the atmosphere. \textit{Annu.
  Rev. Earth Planet. Sci.}, \textbf{34}, 655--688.

\bibitem[{Schneider and Bordoni(2008)Schneider, and
  Bordoni}]{schneider_eddy-mediated_2008}
Schneider, T., and S.~Bordoni, 2008: Eddy-{{Mediated Regime Transitions}} in
  the {{Seasonal Cycle}} of a {{Hadley Circulation}} and {{Implications}} for
  {{Monsoon Dynamics}}. \textit{Journal of the Atmospheric Sciences},
  \textbf{65~(3)}, 915--934, \doi{10.1175/2007JAS2415.1}.

\bibitem[{Seo et~al.(2023)Seo, Yoon, Lu, Hu, Staten,, and
  Frierson}]{seo_what_2023}
Seo, K.-H., S.-P. Yoon, J.~Lu, Y.~Hu, P.~W. Staten, and D.~M.~W. Frierson,
  2023: What controls the interannual variation of {{Hadley}} cell extent in
  the {{Northern Hemisphere}}: Physical mechanism and empirical model for edge
  variation. \textit{npj Climate and Atmospheric Science}, \textbf{6~(1)},
  1--12, \doi{10.1038/s41612-023-00533-w}.

\bibitem[{Showman et~al.(2014)Showman, Wordsworth, Merlis,, and
  Kaspi}]{showman_atmospheric_2014}
Showman, A.~P., R.~D. Wordsworth, T.~M. Merlis, and Y.~Kaspi, 2014: Atmospheric
  circulation of terrestrial exoplanets. \textit{Comparative {{Climatology}} of
  {{Terrestrial Planets}}}, Space {{Science}}, 2nd ed., University of Arizona
  Press.

\bibitem[{Singh(2019)}]{singh_limits_2019}
Singh, M.~S., 2019: Limits on the extent of the solsticial {{Hadley Cell}}:
  {{The}} role of planetary rotation. \textit{Journal of the Atmospheric
  Sciences}, \textbf{76~(7)}, 1989--2004, \doi{10.1175/JAS-D-18-0341.1}.

\bibitem[{Sobel and Schneider(2009)Sobel, and
  Schneider}]{sobel_single-layer_2009}
Sobel, A.~H., and T.~Schneider, 2009: Single-layer axisymmetric model for a
  {{Hadley}} circulation with parameterized eddy momentum forcing.
  \textit{Journal of Advances in Modeling Earth Systems}, \textbf{1~(3)}, 10,
  \doi{10.3894/JAMES.2009.1.10}.

\bibitem[{Staten et~al.(2020)}]{staten_tropical_2020}
Staten, P.~W., and Coauthors, 2020: Tropical widening: {{From}} global
  variations to regional impacts. \textit{Bulletin of the American
  Meteorological Society}, \doi{10.1175/BAMS-D-19-0047.1}.

\bibitem[{Tandon et~al.(2013)Tandon, Gerber, Sobel,, and
  Polvani}]{tandon_understanding_2013}
Tandon, N.~F., E.~P. Gerber, A.~H. Sobel, and L.~M. Polvani, 2013:
  Understanding {{Hadley Cell Expansion}} versus {{Contraction}}: {{Insights}}
  from {{Simplified Models}} and {{Implications}} for {{Recent Observations}}.
  \textit{Journal of Climate}, \textbf{26~(12)}, 4304--4321,
  \doi{10.1175/JCLI-D-12-00598.1}.

\bibitem[{Vallis et~al.(2015)Vallis, {Zurita-Gotor}, Cairns,, and
  Kidston}]{vallis_response_2015}
Vallis, G.~K., P.~{Zurita-Gotor}, C.~Cairns, and J.~Kidston, 2015: Response of
  the large-scale structure of the atmosphere to global warming.
  \textit{Quarterly Journal of the Royal Meteorological Society},
  \textbf{141~(690)}, 1479--1501, \doi{10.1002/qj.2456}.

\bibitem[{Voigt et~al.(2016)}]{voigt_tropical_2016}
Voigt, A., and Coauthors, 2016: The tropical rain belts with an annual cycle
  and a continent model intercomparison project: {{TRACMIP}}. \textit{Journal
  of Advances in Modeling Earth Systems}, \textbf{8~(4)}, 1868--1891,
  \doi{10.1002/2016MS000748}.

\bibitem[{Walker and Schneider(2005)Walker, and
  Schneider}]{walker_response_2005}
Walker, C.~C., and T.~Schneider, 2005: Response of idealized {{Hadley}}
  circulations to seasonally varying heating. \textit{Geophysical Research
  Letters}, \textbf{32~(6)}, L06\,813, \doi{10.1029/2004GL022304}.

\bibitem[{Walker and Schneider(2006)Walker, and Schneider}]{walker_eddy_2006}
Walker, C.~C., and T.~Schneider, 2006: Eddy influences on {{Hadley}}
  circulations: {{Simulations}} with an idealized {{GCM}}. \textit{Journal of
  the Atmospheric Sciences}, \textbf{63~(12)}, 3333--3350,
  \doi{10.1175/JAS3821.1}.

\bibitem[{{Watt-Meyer} and Frierson(2019){Watt-Meyer}, and
  Frierson}]{watt-meyer_itcz_2019}
{Watt-Meyer}, O., and D.~M.~W. Frierson, 2019: {{ITCZ Width Controls}} on
  {{Hadley Cell Extent}} and {{Eddy-Driven Jet Position}} and {{Their
  Response}} to {{Warming}}. \textit{Journal of Climate}, \textbf{32~(4)},
  1151--1166, \doi{10.1175/JCLI-D-18-0434.1}.

\bibitem[{{Watt-Meyer} et~al.(2019){Watt-Meyer}, Frierson,, and
  Fu}]{watt-meyer_hemispheric_2019}
{Watt-Meyer}, O., D.~M.~W. Frierson, and Q.~Fu, 2019: Hemispheric {{Asymmetry}}
  of {{Tropical Expansion Under CO2 Forcing}}. \textit{Geophysical Research
  Letters}, \textbf{46~(15)}, 9231--9240, \doi{10.1029/2019GL083695}.

\bibitem[{Wei and Bordoni(2018)Wei, and Bordoni}]{wei_energetic_2018}
Wei, H.-H., and S.~Bordoni, 2018: Energetic {{Constraints}} on the {{ITCZ
  Position}} in {{Idealized Simulations With}} a {{Seasonal Cycle}}.
  \textit{Journal of Advances in Modeling Earth Systems}, \textbf{10~(7)},
  1708--1725, \doi{10.1029/2018MS001313}.

\bibitem[{WMO(1957)}]{wmo_meteorology_1957}
WMO, 1957: Meteorology --- a three-dimensional science. \textit{WMO Bulletin},
  \textbf{6~(4)}, 134--138.

\bibitem[{Zalucha et~al.(2010)Zalucha, Plumb,, and
  Wilson}]{zalucha_analysis_2010}
Zalucha, A.~M., R.~A. Plumb, and R.~J. Wilson, 2010: An {{Analysis}} of the
  {{Effect}} of {{Topography}} on the {{Martian Hadley Cells}}. \textit{Journal
  of the Atmospheric Sciences}, \textbf{67~(3)}, 673--693,
  \doi{10.1175/2009JAS3130.1}.

\bibitem[{Zhang and Lutsko(2022)Zhang, and Lutsko}]{zhang_seasonal_2022}
Zhang, P., and N.~J. Lutsko, 2022: Seasonal {{Superrotation}} in {{Earth}}'s
  {{Troposphere}}. \textit{Journal of the Atmospheric Sciences},
  \textbf{79~(12)}, 3297--3314, \doi{10.1175/JAS-D-22-0066.1}.

\bibitem[{Zhou and Xie(2018)Zhou, and Xie}]{zhou_hierarchy_2018}
Zhou, W., and S.-P. Xie, 2018: A {{Hierarchy}} of idealized monsoons in an
  intermediate {{GCM}}. \textit{Journal of Climate}, \textbf{31~(22)},
  9021--9036, \doi{10.1175/JCLI-D-18-0084.1}.

\bibitem[{{Zurita-Gotor} and {\'A}lvarez-Zapatero(2018){Zurita-Gotor}, and
  {\'A}lvarez-Zapatero}]{zurita-gotor_coupled_2018}
{Zurita-Gotor}, P., and P.~{\'A}lvarez-Zapatero, 2018: Coupled {{Interannual
  Variability}} of the {{Hadley}} and {{Ferrel Cells}}. \textit{Journal of
  Climate}, \textbf{31~(12)}, 4757--4773, \doi{10.1175/JCLI-D-17-0752.1}.

\end{thebibliography}

\end{document}